\documentclass[iop]{emulateapj}
\usepackage{natbib}
\usepackage{amsmath}
\usepackage{xcolor}
\usepackage{hyperref}
\usepackage{academicons}
\definecolor{orcidlogocol}{HTML}{A6CE39}

\newcommand{\Kepler}{{\sl Kepler}\ }

\newcommand{\be}{\begin{equation}}
\newcommand{\ee}{\end{equation}}

\slugcomment{}
\shorttitle{ Grazing transits }
\shortauthors{Gilbert}

\begin{document}
\title{
Accurate modeling of grazing transits using umbrella sampling
}
\author{
Gregory J. Gilbert\altaffilmark{1}
}
\altaffiltext{1}{Department of Astronomy and Astrophysics, University of Chicago, 5640 S. Ellis Ave., Chicago, IL 60637, USA}
\email{Email: gjgilbert@uchicago.edu}

\begin{abstract}

Grazing transits present a special problem for statistical studies of exoplanets. Even though grazing planetary orbits are rare (due to geometric selection effects), for many low to moderate signal-to-noise cases, a significant fraction of the posterior distribution is nonetheless consistent with a grazing geometry. A failure to accurately model grazing transits can therefore lead to biased inferences even for cases where the planet is not actually on a grazing trajectory. With recent advances in stellar characterization, the limiting factor for many scientific applications is now the quality of available transit fits themselves, and so the time is ripe to revisit the transit fitting problem. In this paper, we model exoplanet transits using a novel application of umbrella sampling and a geometry-dependent parameter basis that minimizes covariances between transit parameters. Our technique splits the transit fitting problem into independent Monte Carlo sampling runs for the grazing, non-grazing, and transition regions of the parameter space, which we then recombine into a single joint posterior probability distribution using a robust weighting scheme. Our method can be trivially parallelized and so requires no increase in the wall clock time needed for computations. Most importantly, our method produces accurate estimates of exoplanet properties for both grazing and non-grazing orbits, yielding more robust results than standard methods for many common star-planet configurations.
\end{abstract}


\section{Introduction}\label{sec:intro}

Roughly 75\% of all known exoplanets have been discovered via transit surveys, giving transiting planets an outsized influence on our ability to constrain both exoplanet demographics and planet formation models. In most cases, our best answers remain data-limited: the precision of available transit measurements is insufficient to distinguish between theoretical models. Until recently, uncertainties on transit parameters were dominated by uncertainties on stellar quantities (especially stellar radii), but with recent advances in stellar characterization via \textit{Gaia} astrometry \citep{GaiaDR2, Berger2018} and via high-quality spectroscopy \citep[e.g.][]{Petigura2017, Johnson2017}, the achievable precision on planetary radii and orbital elements derived from transit surveys is now limited predominantly by the quality of the transit fits themselves \citep{Petigura2020}.

Obtaining higher precision planet parameter estimates will allow us to answer several pressing open questions in exoplanet science. For example, is the so-called radius valley fully or only partially depleted of planets \citep{Fulton2017, FultonPetigura2018, VanEylen2018, HardegreeUllman2020, Petigura2020}? Are patterns in the mass-period-radius distribution sculpted primarily by photoevaporation \citep{Owen2017}, by core-powered mass loss \citep{Ginzburg2018, Gupta2019}, or by some combination of mechanisms \citep{NeilRogers2020}. What do correlations between radii, periods, inclinations, and eccentricities tell us about the processes by which planets form and evolve \citep{Hansen2012, Chiang2013}? Techniques for analyzing exoplanet demographics and exo-system architectures have already been extensively developed \citep{Howard2012, Fabrycky2014, Milholland2017, Weiss2018, He2019, Mills2019, Christiansen2020, GilbertFabrycky2020, HardegreeUllman2020, Zink2020}, so once sufficiently precise estimates of planet properties for a suitably large sample of objects become available, their astrophysical implications will become almost immediately apparent.

The keystone transit parameter which must be accurately estimated in order to allow reliable measurements of all underlying planet properties is the impact parameter. Unfortunately, impact parameters are notoriously difficult to estimate and consequently have seldom been the focus of transit lightcurve analyses. Instead, most previous studies have focused on measuring transit depths and durations, both of which are usually well-constrained by observations \citep[e.g.][]{ Mullally2015, Thompson2018}. Still, the limiting factor when converting transit observables into planetary radii, inclinations, and eccentricities is more often than not the least well constrained variable, so we must address the challenge of measuring impact parameters head-on.

This problem is vital because transit observables can rarely be translated into planet properties on a one-to-one basis. For planetary radii, this condition arises because transit depths contain information about stellar limb darkening in addition to information about planetary radius, while for inclinations and eccentricities the condition arises because transit durations contain information about transit chord length as well as information about orbital velocity. In all cases, accurately deriving planet parameters from transit observations requires knowing how far from the stellar center the planet transits, which is precisely what the impact parameter measures. Even quantities which are usually considered to by well constrained - most notably the planet-to-star radius ratio - depend implicitly (and sometimes sensitively) on the assumed impact parameter. The methods developed in this paper thus first and foremost represent a means of obtaining the highest quality impact parameter measurements possible as a stepping stone toward obtaining correspondingly high quality estimates of planetary radii and other orbital elements.

The degeneracies between transit parameters become most severe for planets on grazing or near-grazing trajectories. In particular, although the planet-to-star radius ratio, $r \equiv r_p/R_{\star}$, and the normalized impact parameter, $b$, are largely uncorrelated for non-grazing orbits, these two parameters become highly correlated for grazing geometries. Because grazing orbits are rare \citep{KippingSandford2016}, the standard approach has been to reject suspected grazing or near-grazing bodies from statistical studies all together \citep[e.g.][]{Petigura2020}. However, for many low to moderate signal-to-noise cases, a significant fraction of the posterior distribution is nonetheless consistent with a grazing geometry, even for planets which most likely orbit on non-grazing trajectories. A failure to accurately model grazing transits can therefore lead to biased inferences even for cases where the planet is not actually on a grazing orbit. Although the $r-b$ grazing degeneracy has been known about for some time \citep{Rowe2014, Rowe2015}, the severity of the problem as it pertains to near-grazing orbits (i.e. orbits with $0.7 \lesssim b < 1-r$) has only recently begun to be appreciated. Because planetary orbits are (probably) isotropically oriented, the unavoidable conclusion is that as many as one third of all existing transit measurements may be corrupted by incomplete consideration of grazing transit geometries. The true situation is probably not so dire, but without a reliable way to distinguish grazing from non-grazing orbits, it is difficult to know which transit measurements should be trusted.

The goal of this paper is twofold: first, to find a way to reliably identify grazing and near-grazing transits, and second, to accurately fit these transits using a method that produces robust estimates of exoplanet properties. The tool for the job is umbrella sampling \citep{TorrieValleau1977}, a statistical technique which is closely related to importance sampling. Although the application of umbrella sampling is standard practice is the field of molecular dynamics where it originated, umbrella sampling methods have only recently begun to be applied to astrophysical problems \citep{Matthews2018}. Umbrella sampling is a powerful tool for sampling multimodal and other complicated posterior distributions and is suitable for many astrophysical applications. So, in addition to addressing the specific problem of transit lightcurve fitting, we aim for this paper to serve as an accessible introduction to umbrella sampling for astronomers unfamiliar with this fruitful technique. Our present work was primarily inspired by \citet{Matthews2018}, which also stemmed from the twin motivations of applying umbrella sampling to an astrophysical problem (in their case, sampling the low-probability tails of distributions in order to compare cosmological models) and also introducing umbrella sampling to astronomers in a pedagogically accessible manner.

This paper is organized as follows. In \S\ref{sec:transits} we review the geometry of transits and discuss the particular challenges of modeling lightcurves of exoplanets on grazing trajectories. In \S\ref{sec:new_basis} we introduce a new parameter basis designed to efficiently sample grazing transits models. In \S\ref{sec:umbrella} we introduce the concept of umbrella sampling and describe our new method for fitting exoplanet transits. In \S\ref{sec:sampler_comparison} we present several case studies illustrating the efficacy of our method in a variety of contexts and demonstrate that umbrella sampling outperforms standard sampling techniques. In \S\ref{sec:real_systems} we apply our method to several real \Kepler Objects of Interest with high impact parameters reported on the NASA exoplanet archive \citep{Akeson2013}.\footnote{\url{https://exoplanetarchive.ipac.caltech.edu}; all data for this work was downloaded 18 July 2021} In \S\ref{sec:summary} we summarize our main results and provide recommendations for future lightcurve modeling efforts.

\section{The geometry of grazing transits}\label{sec:transits}

In this section, we review a few salient aspects of transit lightcurve modeling relevant to the analysis of grazing transits. Exoplanet experts will likely be familiar with much of \S 2.1, which we include in order to aid other astronomers who may wish to adopt umbrella sampling for other problems. A full pedagogical introduction to the geometry of transit lightcurves is presented in \citet{Winn2010}.

\subsection{The transit model}

\begin{figure}
    \centering
    \includegraphics[width=0.45\textwidth]{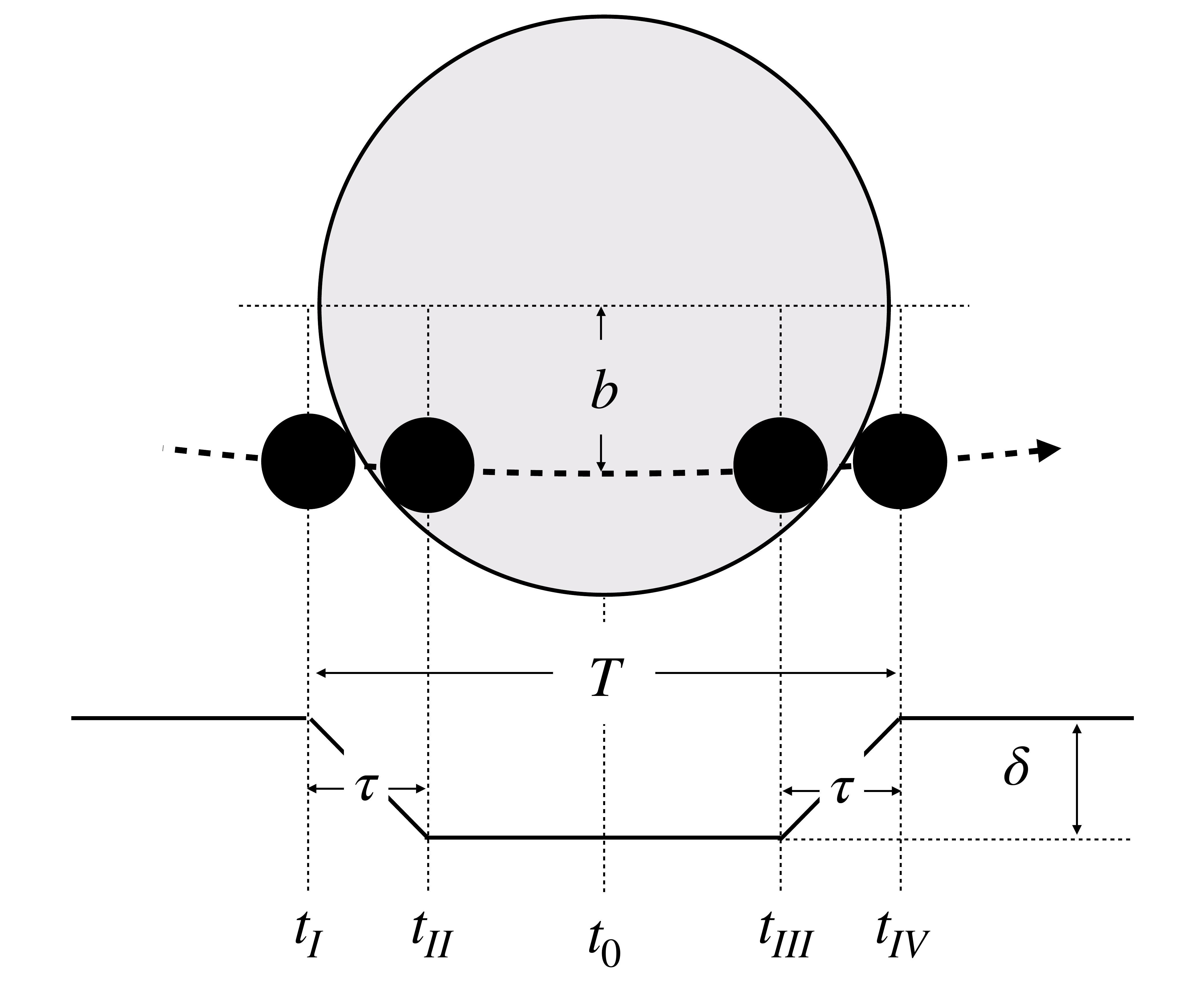}
    \caption{Schematic illustration of a non-grazing transit geometry with the corresponding lightcurve approximated as a trapezoid. The transit depth, $\delta$; transit duration, $T$; ingress/egress duration $\tau$; mid-transit time, $t_0$; impact parameter, $b$; and $1^{\rm st}$ through $4^{\rm th}$ contact points, $t_I$ - $t_{IV}$, are indicated. Note that for this study, the transit duration, $T$, will always refer to the full first-to-fourth contact duration, $T_{14}$, unless otherwise specified because this is the only duration which is defined for all grazing and non-grazing geometries. The approximation $\tau_{12} = \tau_{34}$ (i.e. ingress and egress timescales are equal) is valid as long as eccentricities are not very large. Figure adapted from \citet{Winn2010}.\smallskip}
    \label{fig:transit_geometry}
\end{figure}

The observables which can be directly recovered from a single transit light curve are the mid-transit time, $t_0$, the transit depth, $\delta$, the transit duration, $T$, and the ingress/egress timescale, $\tau$ (Figure \ref{fig:transit_geometry}). When multiple transits are observed, the orbital period, $P$, can be inferred as well. If the planet is assumed to be on a circular orbit around an isolated, uniform surface brightness host star, one can immediately derive four physical quantities: the planet-to-star radius ratio, $r \equiv r_p/R_{\star}$, the scaled separation, $a/R_{\star}$, the mean stellar density, $\rho_{\star}$, and the normalized impact parameter, $b$. 

The transit lightcurve (again, assuming a circular orbit and a solitary star) can be can be fully specified by any non-degenerate combination using five of the nine above parameters, or any derivable quantity thereof \citep{SeagerMallenOrnelas2003, Winn2010}. These five parameters constitute the model \textit{basis set}. In practice, $P$, $t_0$, and $\delta$, are usually well constrained by the data, so most reasonable basis sets will include $P$, $t_0$, and either $\delta$ or $r$ (or slight modifications of these quantities) as the first three free parameters. The transit duration, $T$, is usually well constrained by the data as well, but for even modestly noisy lightcurves, the ingress/egress timescale, $\tau$, is difficult to resolve, making the selection of the final basis parameter a non-trivial task. Moreover, because $\tau \approx 5$ min is shorter than the observing cadence in many cases ($\Delta t \approx 30$ min for Kepler long cadence data and TESS primary mission full frame images), data binning often precludes a precise ingress/egress characterization even for high signal-to-noise transits \citep{Kipping2010, PriceRogers2014}. The ratio $T/\tau$ is as important as the values of $T$ and $\tau$ in isolation, so the choices of these final two basis parameters must be considered in tandem.

Throughout this study, $T$ will refer to the full first-to-fourth contact transit duration $T_{14}$ unless otherwise noted because this is the only transit duration which is readily defined for all grazing and non-grazing geometries. In \S\ref{sec:summary} we will discuss how to modify our model to incorporate different transit durations which may be better constrained by the data. For the time being, limiting consideration to $T_{14}$ simplifies our discussion considerably and allows us to focus on the ideas which are unique to this paper.

A straightforward approach to selecting the final basis pair is to use $\{T, \tau\}$ directly, although one might just as easily choose $\{b, \rho_{\star}\}$, $\{T_{14}, T_{23}\}$ or even more exotic pairs such as $\{b^2, 1/T\}$. Numerous basis sets have been proposed in the literature \citep[e.g.][]{Bakos2007, Carter2008, Pal2008, Eastman2013}, although none has yet been adopted as the standard set applied in all cases. This lack of a standard basis set and the wide variety of different parameterizations in use speaks to the subtle challenge of transit model fitting. For the present work, we will primarily use the basis pair $\{T, b\}$ as our final two model parameters.

To account for nonzero stellar limb darkening, we adopt the standard approach and employ a quadratic limb darkening profile \citep{Claret2000, MandelAgol2002} using the efficient $\{q_1, q_2\}$ parameterization introduced by \citet{Kipping2013}.

To account for nonzero eccentricity, we employ the photoeccentric effect \citep{FordQuinnVeras2008, DawsonJohnson2012} to compare the stellar density implied by a circular transit model, $\rho_{\rm circ}$, to an independent measurement (e.g. from spectroscopy or asteroseismology) of the stellar density, $\rho_{\rm obs}$, via the relation

\begin{equation}\label{eq:photoeccentric}
    \frac{\rho_{\rm circ}}{\rho_{\rm obs}} = \Big( \frac{1 + e\sin\omega}{\sqrt{1-e^2}} \Big)^3
\end{equation}

where $e$ is the eccentricity and $\omega$ is the longitude of periastron. The main advantage of this indirect method over the straightforward approach - i.e. fitting $e$ and $\omega$ directly - is that circular orbits are both faster to compute and require two fewer variables to describe than do non-circular orbits.

With all of this in mind, we will take as our fiducial basis set the parameters $\{P, t_0, \ln r, b, \ln T, q_1, q_2\}$, where the logarithms on $r$ and $T$ enforces positivity of the two scale parameters and facilitates sampling over multiple orders of magnitude. As we will see shortly, this basis set performs quite well for non-grazing transits but performs poorly for grazing transits due to an emergent degeneracy between $r$ and $b$ in the grazing regime.

\subsection{Model degeneracy in the grazing regime}\label{subsec:degeneracy}

We will now consider the specific challenges that arise when modeling transit lightcurves of exoplanets on grazing trajectories. In the discussion that follows, the term ``grazing'' refers to any transit geometry for which the planetary disk does not fully overlap the stellar disk at the mid-transit point. In other words, we consider a transit to be non-grazing if $b \leq 1 - r$ and grazing if $b > 1 - r$.

\begin{figure}
    \centering
    \includegraphics[width=0.45\textwidth]{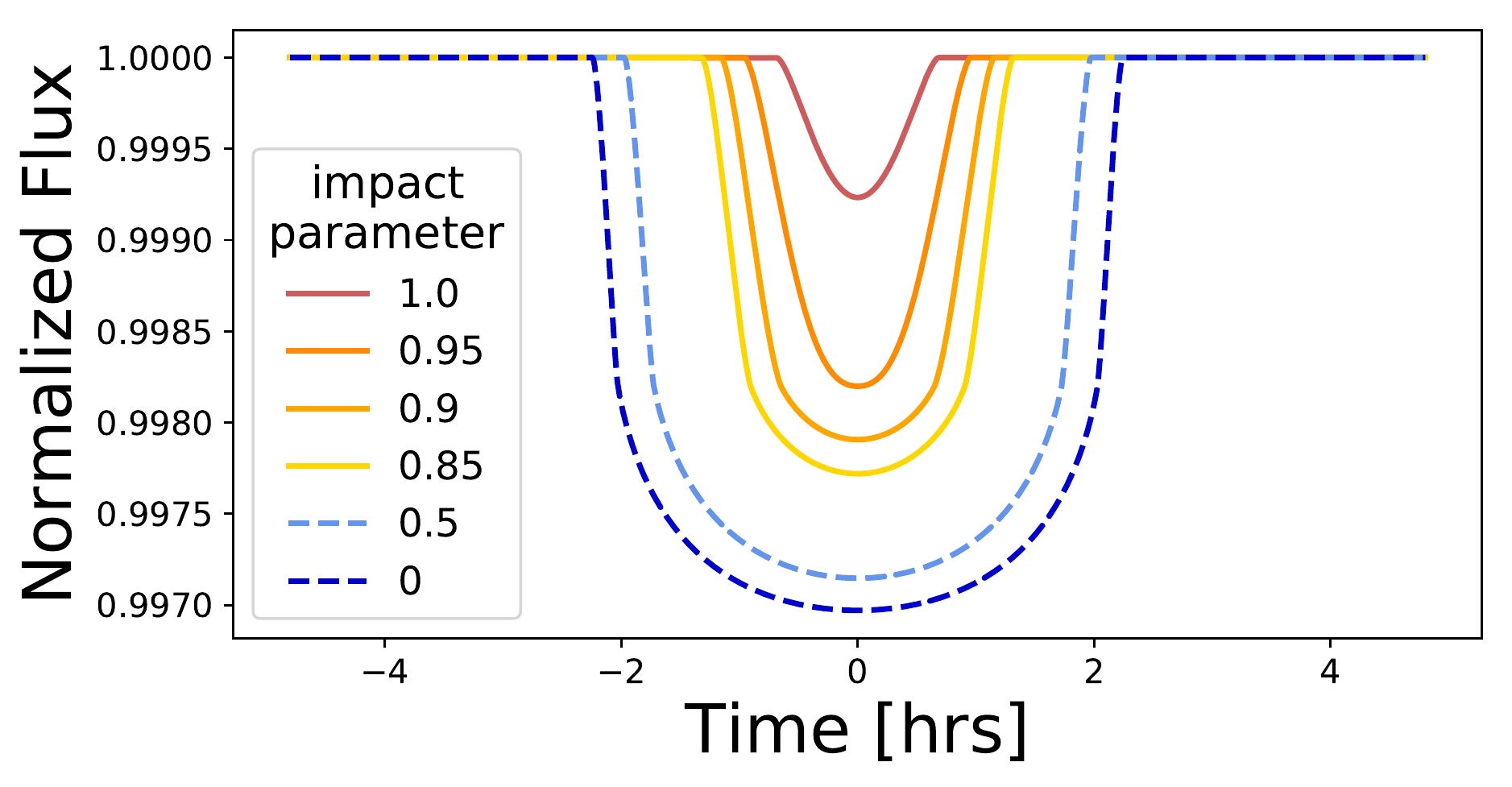}
    \caption{Lightcurve model illustrating how the transit shape changes as a function on impact parameter, $b$. This model is for a warm Neptune orbiting a Sun-like star with model parameters $r=0.05$, $P=13.0$ days, $R_{\star} = R_{\odot}$, $M_{\star} = M_{\odot}$, $u=(0.4,0.25)$. Warm colored, solid lines indicate grazing or near-grazing transits (informally defined here as $b \gtrsim 0.8$), while cool-colored, dashed lines indicate non-grazing transits. As $b$ increases, transit duration decreases (due to the shorter transit chord) and transit depth decreases (due to stellar limb darkening). There is little change in the transit shape between $0 \leq b \lesssim 0.5$, making differences between low impact parameters difficult to resolve. The transit depth and duration both change more rapidly above $b \gtrsim 0.5$. At $b = 1 - r = 0.95$, the transit morphology switches from ``U-shaped'' to ``V-shaped'', providing a diagnostic avenue for distinguishing grazing from non-grazing transits, although the effects of limb darkening blur this transition somewhat.}
    \label{fig:grazing_transit_shape}
\end{figure}

As a planet's trajectory moves from low-$b$ to high-$b$, both the transit duration, $T$, and transit depth, $\delta$, are reduced (Figure \ref{fig:grazing_transit_shape}). The reduction in $T$ occurs because the transit chord is shortened, and the reduction in $\delta$ occurs because at high $b$ the planet crosses a dimmer region of the limb-darkened stellar disk. When the planet crosses the grazing boundary at $b = 1-r$, the transit switches from U-shaped to V-shaped. For high signal-to-noise cases, this change in morphology can be used to distinguish between grazing and non-grazing transits, but for lower signal-to-noise cases, there is enough model flexibility that the transit shape - and by extension the transit geometry - remains ambiguous. 

This ambiguity means that it is often necessary to sample from the grazing regime even for planets which are not actually on grazing trajectories. Unfortunately, two complications arise when attempting to simultaneously sample from both the grazing and non-grazing regions of the posterior distribution. First, although our fiducial basis set performs well when $b < 1-r$, for grazing trajectories $r$ and $b$ become highly correlated, producing a narrow degeneracy ridge that is difficult to explore (more on this in a moment). Second, because the posterior topology is extremely different on either side of the grazing transition, a ``bottleneck'' or ``funnel'' arises and the sampler often struggles to cross this threshold.

In the grazing regime, as a planet moves to higher $b$ with fixed $r$, the overlap area between the stellar and planetary disk will decrease, thereby reducing the transit depth. However, if $r$ is allowed to float as a free parameter, a large $b$ can be compensated for by a commensurate increase in $r$. Thus $r$ and $b$ become almost perfectly positively correlated and sometimes a sampler will find an extremely large radius ($r_p \gg R_{\star}$) and extremely high impact parameter ($b \gg 1$), which is obviously unphysical. This is a well known problem in transit fitting \citep{Rowe2014, Rowe2015} and is a clear case where common sense is in conflict with the analysis.

\begin{figure*}
    \centering
    \includegraphics[width=0.9\textwidth]{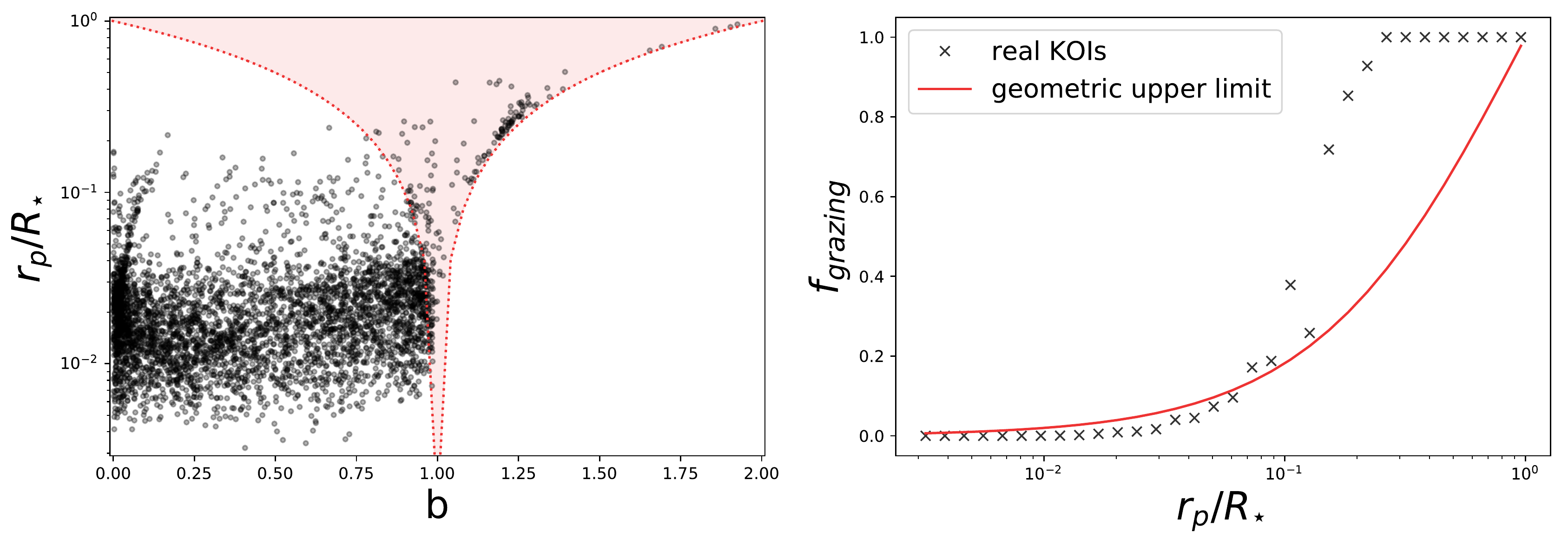}
    \caption{Distribution of $r_p/R_{\star}$ and $b$ for cumulative Kepler Object of Interest (KOI) planet candidates. \textit{Left panel}: joint 2D distribution of $r_p/R_{\star}$ and $b$. Each point represents an individual validated or candidate planet. The red shaded region highlights planets inferred to orbit on a grazing trajectory. There is a suspicious pile-up of planets near $b\approx1.25$ at the $b=1-r$ boundary, hinting that the radius and impact parameter measurements derived for these planets may not be reliable. Non-isotropic structure in the distribution among planets on non-grazing trajectories - particularly near $b=0$ - suggests that measurements for these planets should be approached with some skepticism as well. \textit{Right panel}: fraction of planets inferred to be on grazing trajectories as a function of radius ratio. The red line plots the relation $f=2r/(1+r)$, which is the geometric upper limit on how many planets are expected to be on grazing orbits, ignoring any reduced detection efficiencies for grazing transits. For planets with physically plausible radii ($r_p \lesssim 2 R_J$), the observed fraction of grazing transits is in line with expectations, but for super-giant planets an overabundance of KOIs are found on grazing trajectories, again suggesting that their radius and impact parameter measurements may be unreliable.}
    \label{fig:koi_supergiants}
\end{figure*}

The effects of the $r-b$ degeneracy can be readily seen upon inspection of real Kepler data (Figure \ref{fig:koi_supergiants}). Compared to isotropic expectations for the cumulative Kepler Object of Interest (KOI) catalog, there is an overabundance of super-giant planets ($r_p \gtrsim 2 R_J$) found on grazing trajectories, but for objects with astrophysically sensible radii, the fraction of planets inferred to be on grazing trajectories is roughly in line with expectations. Furthermore, nearly every implausibly large super-giant planet has a quoted impact parameter consistent with a grazing trajectory, and many of these suspicious objects cluster at the $b=1+r$ boundary that marks where planets are not only on grazing orbits, but on \textit{extremely} grazing orbits for which the planetary and stellar disks barely overlap at all. While it is possible there is some complicated selection effect at play wherein only super-giants on grazing orbits pass all vetting thresholds necessary to be included in the KOI database, the simpler explanation is that the majority of these supposed super-giants are actually super-Earths or mini-Neptunes on non-grazing orbits, with inferred $r$ and $b$ values that are artifacts of a transit fitting procedure gone awry.

The effects of the transition threshold bottleneck are more difficult to notice because the most common outcome is that the sampler fails to enter the grazing regime entirely. In this situation, an individual Markov chain may appear well mixed, even though the posterior distribution has not been fully explored. This scenario is arguably worse than when the sampler becomes stuck deep in the grazing regime at astrophysically implausible values of $r$ because we often will not recognize that anything has gone wrong, even after consulting the usual set of Markov chain diagnostics. If we cannot trust that our sampler is well-behaved near the grazing/non-grazing boundary, we cannot trust that our sampler is well-behaved anywhere. We must therefore be skeptical of any results obtained before we can be confident that the bottleneck has not biased our inferences.

\begin{figure}
    \centering
    \includegraphics[width=0.45\textwidth]{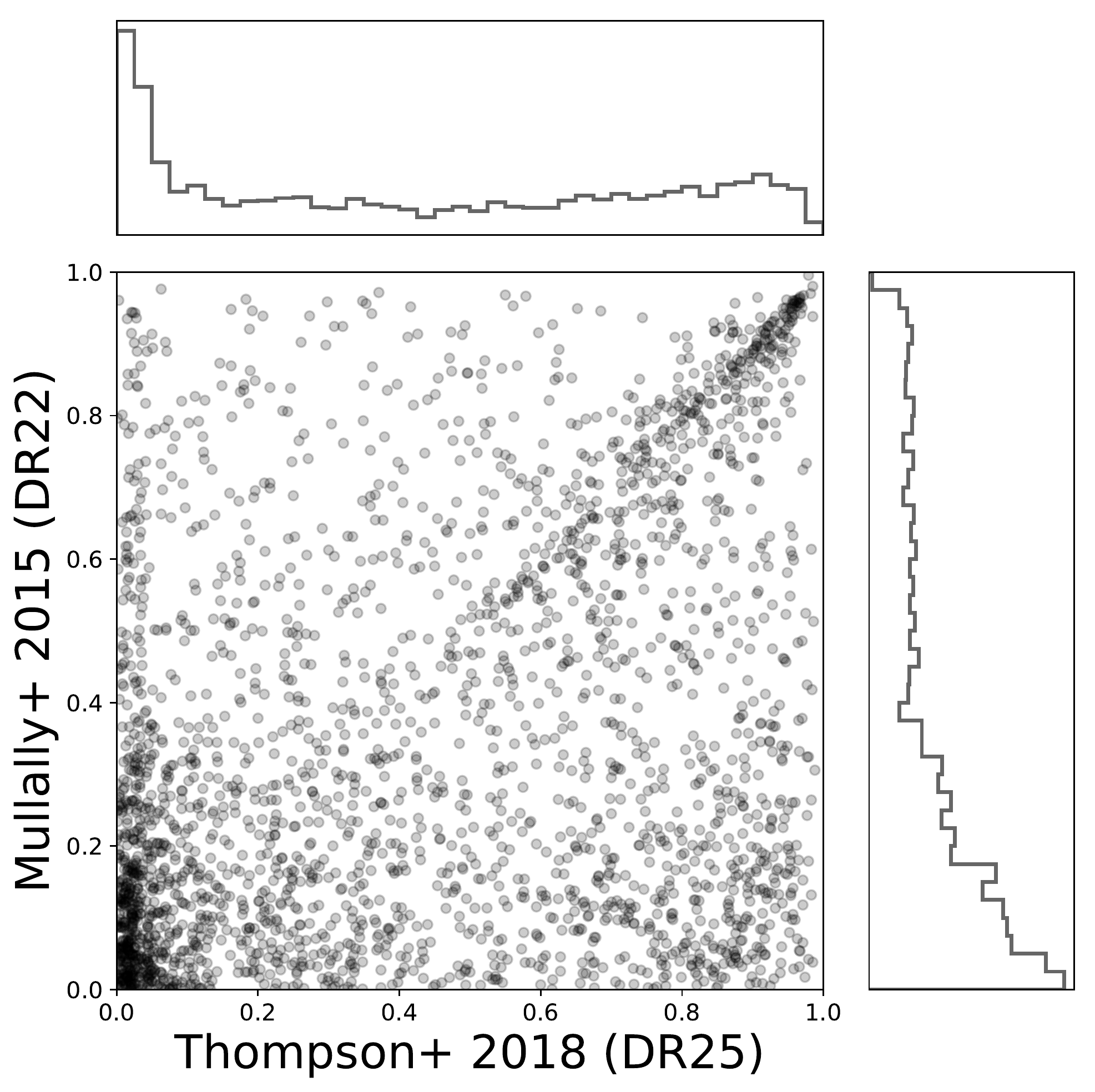}
    \caption{Impact parameters reported by Kepler DR22 \citep{Mullally2015} vs DR25 \citep{Thompson2018}. Plotted values correspond to best-fit point estimates. Each point indicates a single Kepler Object of Interest reported in both catalogs, so points should cluster around the line $b_{22}=b_{25}$. The high degree of scatter in the actual data indicates that results are inconsistent and therefore unreliable. While there is some correspondence of values where $b \gtrsim 0.7$, in a substantial fraction of cases one catalog reports $b\approx0$ while the other reports $b\approx1$. A pile-up of reported values near $b=0$ can be seen in both catalogs, indicating that results are inaccurate. Because $r$ and $b$ are correlated for stars with non-negligible limb darkening, any mismeasurement of $b$ will propagate through to a mismeasurement of $r$.}
    \label{fig:dr22_vs_dr25}
\end{figure}

Comparing impact parameter measurements between different Kepler data releases bears out our exhortation toward caution. Each point in Figure \ref{fig:dr22_vs_dr25} marks an individual KOI reported in both DR22 \citep{Mullally2015} and DR25 \citep{Thompson2018}, so points should cluster around the line $b_{22} = b_{25}$. The high degree of scatter indicates that the actual results are inconsistent, despite the fact that they were obtained from nearly identical input observations and similar data processing pipelines. Although there is some evident correspondence between the catalogs for high impact parameters ($b \gtrsim 0.7$), there is nonetheless a substantial fraction of points which report $b \approx 0$ in one catalog and $b \approx 1$ in the other. Even the 1D single-catalog distributions exhibit inhomogeneity, with a pile-up of reported impact parameters near $b=0$ seen in both catalogs. Although some of these discrepancies can be mitigated by using posterior medians rather than the default maximum likelihood point estimates \citep{Petigura2020}, much of the error is endemic to the problem of impact parameter measurement.

\subsection{Experiments using simulated data}\label{subsec:simulation_description}

In order to illuminate the origin of the skewed $r-b$ distribution seen in real data, we perform an experiment which applies a Markov Chain Monte Carlo (MCMC) model fitting routine to synthetic data. In the next several paragraphs, we describe our method for simulating data and for subsequently fitting a transit model to that data using Hamiltonian Monte Carlo \citep[HMC;][]{Neal2011}. The casual reader may wish to skim these paragraphs so as not to become bogged down in the details. The important point is that we simulate an ordinary transit of an unremarkable star-planet system and then model that transit using our fiducial basis set and standard Monte Carlo sampling techniques.

\begin{figure}
    \centering
    \includegraphics[width=0.45\textwidth]{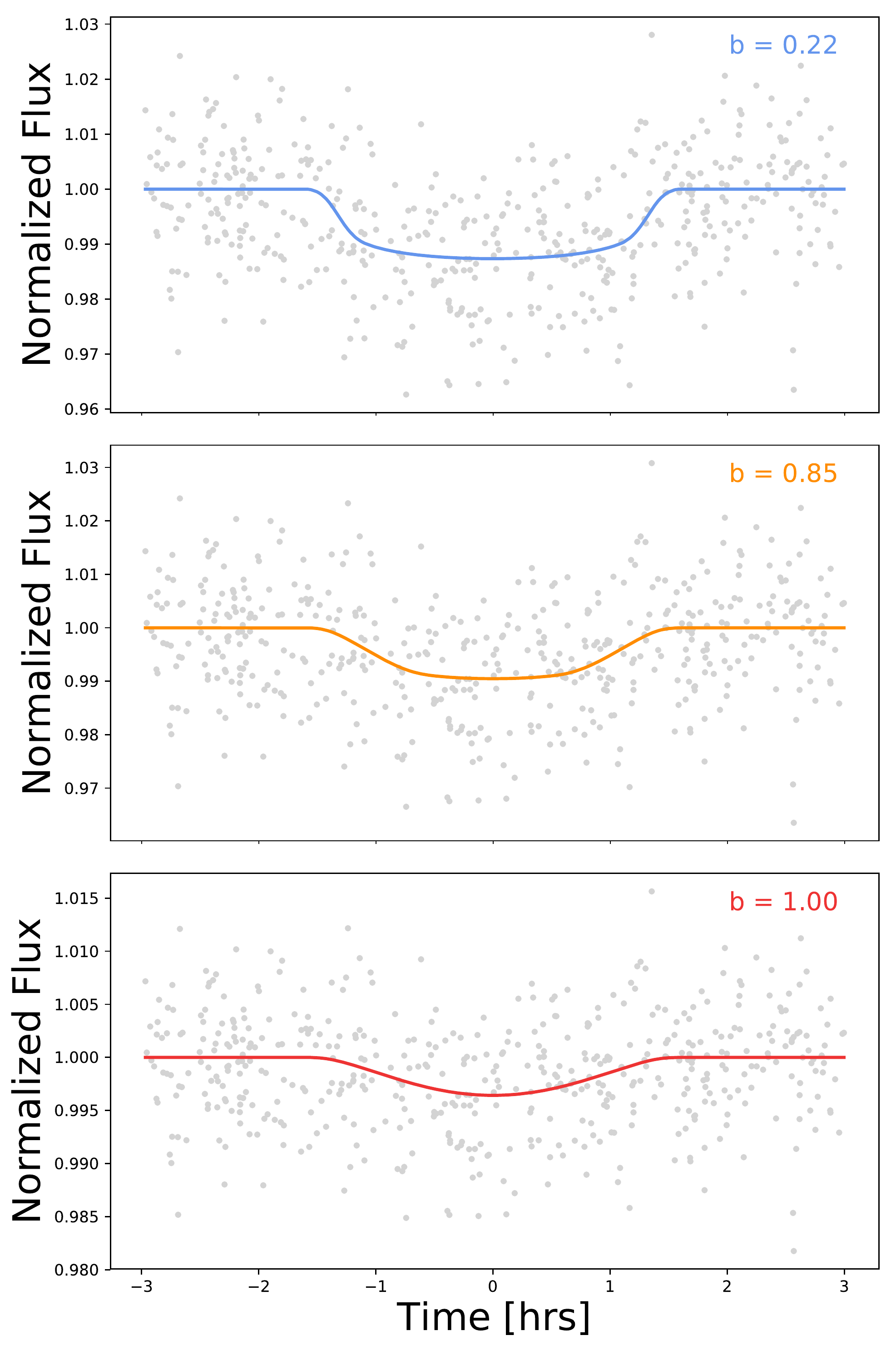}
    \caption{Simulated lightcurve photometry for a Jupiter-size planet on a circular orbit around a Sun-like star with various impact parameters. The orbital period was adjusted to maintain a consistent transit duration of $T=3$ hrs. The solid colored lines show the true underlying model while the grey points have additive Gaussian noise. The low signal-to-noise ratio of the transit makes the orbital trajectory (grazing vs. non-grazing) ambiguous. \textit{Top panel}: $b=0.22$, placing the planet on a non-grazing trajectory, corresponding to model J-22 in Table \ref{tab:sim_parameters}. \textit{Middle panel}: $b=0.85$, a near-grazing trajectory (model J-85). \textit{Bottom panel}: $b=1.0$, a grazing trajectory (model J-100).}
    \label{fig:simulated_photometry}
\end{figure}

 The details of our data simulation procedure are as follows. First, we simulate a low signal-to-noise transit of a warm Jupiter on a near-grazing, $P=13$ day circular orbit around a Sun-like star. The period was calculated in order to create a 3 hr transit duration for an impact parameter $b=0.85$. We generated 500 data points between $t_0 \pm T$, each with an integrated exposure time of 14.4 minutes (0.01 days). The finite data points were spaced randomly over the interval in order to minimize aliasing artifacts that might arise from a uniform observing cadence. We then added $\sigma_F/F = 10^4$ ppm Gaussian noise to the data. We did not include any long term trends or correlated noise in our simulation. Ground-truth parameter values are collected as simulation J-85 in Table \ref{tab:sim_parameters}, and the simulated photometry is shown in the middle panel of Figure \ref{fig:simulated_photometry}.

 \begin{table*}
    \centering
    \begin{tabular}{c c | c c c | c c | c}
         Parameter &  Unit & J-22 & J-85 & J-100 & SE & MN & MHZ\\
         \hline
         \textbf{Star} \\
         $R_{\star}$ & $R_{\odot}$ & 1.0 & 1.0 & 1.0 & 0.92 & 0.92 & 0.37 \\
         $M_{\star}$ & $M_{\odot}$ & 1.0 & 1.0 & 1.0 & 0.86 & 0.86 & 0.38 \\
         $u_1$ & - & 0.40 & 0.40 & 0.40 & 0.48 & 0.48 & 0.46 \\
         $u_2$ & - & 0.25 & 0.25 & 0.25 & 0.22 & 0.22 & 0.28\\
         $\sigma_F$ & ppm & $1\times10^4$ & $1\times10^4$ & $5\times10^3$ & 300 & 300 & 200\\
         
         \textbf{Planet} \\
         $P$ &  days & 3.6 & 13.0 & 44.9 & 21.0 & 21.0 & 37.0 \\
         $r_p$ & $R_{\oplus}$ & 11.2 & 11.2 & 11.2 & 1.3 & 2.2 & 0.38\\
         $b$ & - & 0.22 & 0.85 & 1.00 & 0.70 & 0.98 & 0.70 \\
         $T$ & hrs & 3.0 & 3.0 & 3.0 & 3.24 & 1.26 & 2.33 \\
         \textbf{Derived} \\
         $r$ & - & 0.103 & 0.103 & 0.103 & 0.012 & 0.020 & 0.009 \\
         $\gamma$ & - & 7.57 & 1.36 & 0.0 & 25.2 & 2.48 & 32.7 \\
         $\lambda$ & - & 0.091 & 0.026 & 0.011 & 0.004 & 0.001 & 0.003 \\
         
    \end{tabular}
    \caption{Ground-truth parameter values for simulated lightcurves used to compare a standard sampling approach to our new method. The quantities $\lambda$ and $\gamma$ are defined in Equation \ref{eq:lam_gam}. All simulated planets were placed on circular orbits. The first set of simulations (J-22, J-85, \& J-100) placed a Jupiter-sized planet around a Sun-like star at three different impact parameters in order to produce a non-grazing ($b=0.22$), nearly grazing ($b=0.85)$, and grazing ($b=1.00$) geometry; the orbital period was scaled to preserve a circular orbit for a consistent transit duration $T=3$ hrs. The second pair of simulations (SE, MN), placed a super-Earth ($r_p = 1.6 R_{\oplus}$) and a mini-Neptune ($r_p = 2.2 R_{\oplus}$ on a 21 day orbit around a K star; this experiment was designed to mimic the detection of radius valley planet that will require a precise impact parameter measurement in order to determine the its composition. The final simulation (MHZ) placed a small rocky planet in the habitable zone of an M-dwarf. The properties of the K star were chosen to be similar to to Kepler 20 \citep{Mathur2017}, and the properties the M star were chosen to be similar to GJ 876 \citep{vonBraun2014}. Limb darkening coefficients for all stars were calculated assuming solar metallicity and using the EXOFAST web applet \citep{Eastman2013}. Host star properties are meant to be illustrative of a few different common stellar types rather than an exact match to any real particular stars.}
    \label{tab:sim_parameters}
\end{table*}

We parameterized the model using our fiducial basis set - $\{P, t_0, \ln r, b, \ln T, q_1, q_2\}$ - plus a baseline flux offset, $F_0$, and a white noise jitter term, $\ln\sigma_F$. We held $P$ fixed at the true value and placed uninformative priors on all other variables, the mathematical details of which are collected in Table \ref{tab:priors}. Fixing $P$ is equivalent to assuming that the planet's ephemeris is tightly constrained, which is often the case even for noisy transits. Although in practice most applications will use the best available stellar characterization to place at least a modestly informative prior on $\rho_{\star}$ (and, indirectly, on $T$ and $e$ via the photoeccentric effect), for our present experiment we are more concerned with the sampler behavior (i.e. whether MCMC chains are well behaved) rather than the posterior inferences. Our philosophy is that the model should converge regardless of any particular choice of prior, so we adopt minimally restrictive priors wherever possible.
  
\begin{table}[]
    \centering
    \begin{tabular}{l|l}
         Parameter &  Prior \\
         \hline
         $P$ & \textit{fixed} \\
         $t_0$ & $\mathcal{N}(0.0, 0.1)$ \\
         $\ln r$ & $\mathcal{U}(-9.2,-0.01)$ \\
         $b$ & $\mathcal{U}(1-r,1+r)$ \\
         $\ln T$ & $\mathcal{U}(-4.6,-1.4)$ \\
         $q_1, q_2$ & $\mathcal{U}(0,1)$ \\
         \hline
         $F_0$ & $\mathcal{N}(0,1)$ \\
         $\ln\sigma_F$ & $\mathcal{N}(0,1)$ \\
         \hline
         $\rho_{\star}$ & \textit{see text} \\
         $e$ & \textit{see text}
    \end{tabular}
    \caption{Model priors for the grazing regime. All times are in units of days. $\mathcal{N} = \mathcal{N}(\mu,\sigma)$ denotes normal distributions and $\mathcal{U} = \mathcal{U}(x_{min},x_{max})$ denotes uniform distributions. All units of time are in days or log(days), where appropriate. Note that $b$ is defined as a conditional distribution predicated on $r$, i.e. $p(b) \equiv p(b|r)$; see Appendix \ref{appx:B} for details. Our limb darkening treatment follows \citet{Kipping2013} by placing uninformative priors on the two quadratic coefficients.}
    \label{tab:priors}
\end{table}

We sampled from the posterior distribution using HMC as implemented by \texttt{PyMC3} \citep{pymc3:2016} and the No U-Turn Sampler \citep[NUTS;][]{Hoffman2011}. Each sampling run consisted of two independent chains tuned for 5000 steps and sampled for 1000 draws, for a total of 2000 samples per run. We deliberately left the independent chains short in order to highlight the stochastic nature of the problem, but note that with HMC the autocorrelation length is typically much shorter than for standard random walk Metropolis-Hastings algorithms \citep{Metropolis1953, Hastings1970}, so that the number of effective samples is usually $\gtrsim 25\%$ and under ideal circumstances can approach $100\%$. This high effective sample rate is achievable with HMC because the algorithm adds a ``momentum'' term to the proposal generation process which enables much larger steps sizes than a random walk. During the tuning phase (analogous to the burn-in phase of other MCMC routines), the sampler ``learns'' the posterior topology and adaptively selects an optimal steps size for efficient exploration of the posterior. While the computational cost per step is higher for HMC compared to random walk Metropolis-Hastings, the cost per effective sample is usually considerably lower, especially for high dimensional problems. HMC has only recently begun to gain popularity among astrophysicists, so we direct the interested reader to the excellent review by \citet{Betancourt2017}, as well as tutorials for the Python software packages \texttt{PyMC3}\footnote{\url{https://docs.pymc.io}} \citep{pymc3:2016} and \texttt{exoplanet}\footnote{\url{https://docs.exoplanet.codes}} \citep{ForemanMackey2021}.

\begin{figure*}
    \centering
    \includegraphics[width=0.9\textwidth]{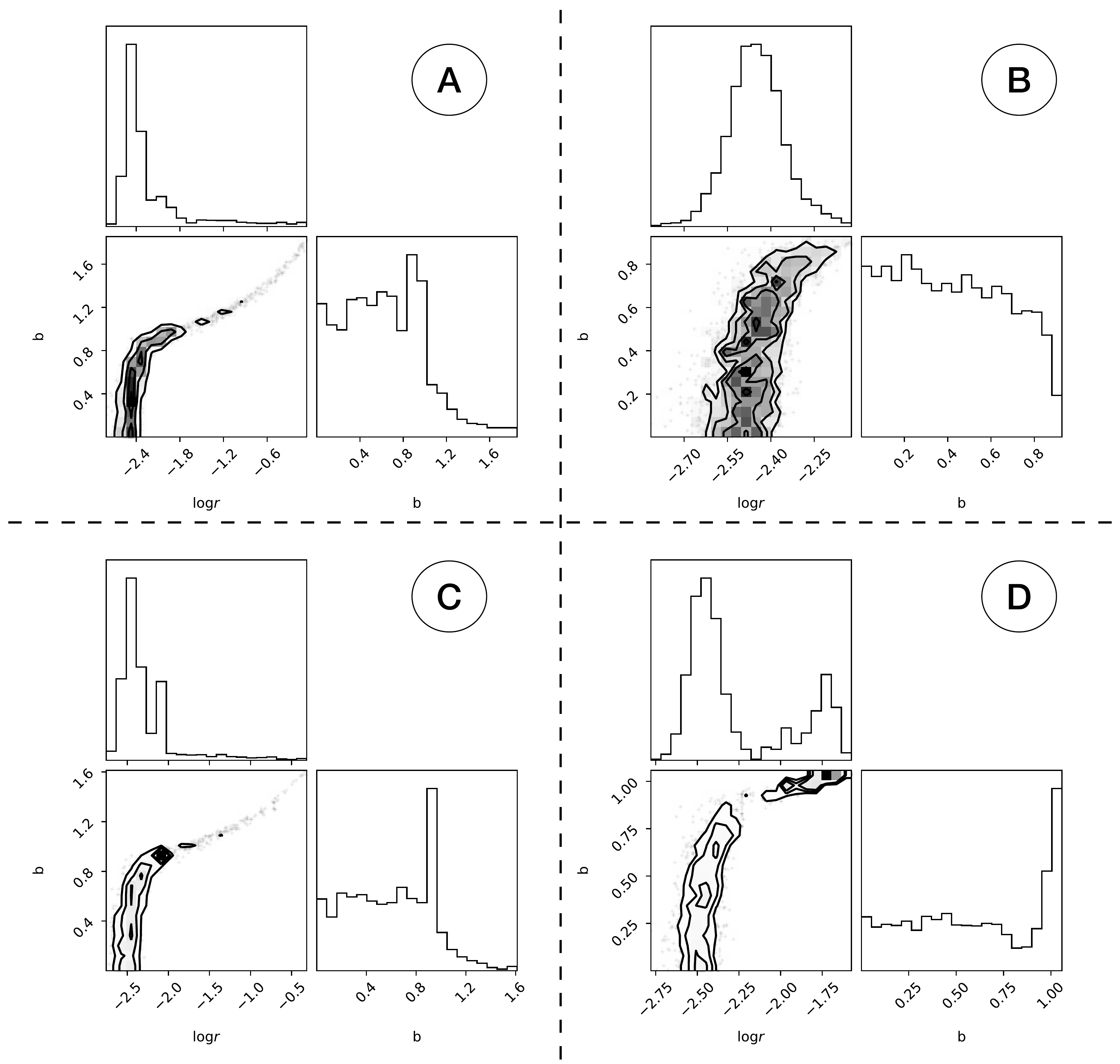}
    \caption{Corner plots of the posteriors from four identical MCMC runs of model J-85 using our fiducial $\{\ln r, b\}$ basis. See \S\ref{subsec:basis_comparison} and Table \ref{tab:sim_parameters} for details of the model setup. The only difference between the runs was the random seed for the Markov chains. Despite their identical setups, each run produces a remarkably different posterior geometry. \textit{Panel A}: The sampler appears to fully explore the posterior region, with most samples consistent with a non-grazing geometry and a smaller fraction extending into the grazing regime. There is a ``dog leg'' feature at $b\approx1$ where the geometry transitions from non-grazing to grazing, and there is a strong degeneracy between $\ln r$ and $b$ for grazing transits. \textit{Panel B}: The sampler fails to explore the grazing regime entirely, giving the illusion of a well-behaved posterior. \textit{Panel C}: The sampler extends to high impact parameters, but catches at the boundary between grazing and non-grazing geometries, producing a sharp spike at $b \approx 1$. \textit{Panel D}: The samples pile up at $b \approx 1$, leading to a bimodal posterior distribution that barely explores the grazing regime at all. Increasing the length the tuning phase and/or the sampling phase does not reliably fix these issues.} 
    \label{fig:rb_corner_degeneracy}
\end{figure*}

Figure \ref{fig:rb_corner_degeneracy} illustrates results of four independent attempts to model simulated transit data using HMC. The only difference from run-to-run was the random seed for the sampler. Despite identical setups, each run produced a remarkably different posterior distribution, sometimes getting stuck in the grazing regime and sometimes failing to explore that regime altogether. The issue is not merely that the chains had not converged, and even increasing the length of the sampling and/or tuning phase by orders of magnitude did not reliably produce consistent results. Because standard sampling methods cannot be counted on to adequately explore both the grazing and non-grazing portions of the distribution, our inferences are unreliable, and we must find a new method for modeling exoplanet transit lightcurves.

The problem is two-fold. First, we need to use a different basis set for grazing vs. non-grazing geometries because the covariance properties of $r$ and $b$ are quite different between the two regimes. Second, we need to find a way to efficiently explore the full posterior space without getting stuck at the grazing transition boundary. The solutions to these problems are interrelated and are discussed in the next two sections of this paper.

\section{A new basis for grazing transits}\label{sec:new_basis}

Our solution to the grazing transit problem is to split the Monte Carlo sampling routine into separate runs for the grazing ($b > 1-r$) and non-grazing  ($b < 1-r$) regimes. We then combine these independent runs into a single posterior distribution using umbrella sampling (see \S\ref{sec:umbrella}). Before we describe our full umbrella sampling procedure, we first present a new basis set which is designed for optimal performance in the grazing regime.

\subsection{Specification of the model parameters}\label{subsec:basis_specification}

Of the seven parameters in our fiducial basis set - $\{P, t_0, \ln r, b, \ln T, q_1, q_2\}$ - four can be carried over to our new grazing basis without modification: $P$, $t_0$, $q_1$, and $q_2$. Both $P$ and $t_0$ are generally tightly constrained by the data and are minimally covariant with the other parameters, and the two limb darkening coefficients, $q_1$ and $q_2$ \citep{Kipping2013}, perform well for both grazing and non-grazing orbits. Only $r$, $b$, and $T$ now remain. The transit duration, $T$, is usually well constrained by the data (albeit somewhat less so than $P$ and $t_0$) and is closely related to the eccentricity via the photoeccentric effect; we therefore maintain $\ln T$ as one of our seven basis parameters. With five parameters in common between the fiducial non-grazing basis and our new grazing basis, our reparameterization effort now hinges on a transformation of $r$ and $b$ (which are highly covariant for grazing transits) into a new parameter pair which is more nearly orthogonal for grazing geometries. Rather than producing new parameters wholesale, our strategy is to find some mapping of $\{r,b\} \rightarrow \{x_1,x_2\}$ with the desired orthogonality when $b > 1-r$.

After some experimentation, we identified a suitable pair of quantities, which we define according to the non-linear combination

\begin{equation}\label{eq:lam_gam}
\begin{aligned}
    &\lambda = r^2 + \beta r\\
    &\gamma = \frac{\beta}{r}
\end{aligned}
\end{equation}

where $\beta \equiv 1-b$ is a convenience variable. Because both $r$ and $b$ are unitless, $\lambda$ and $\gamma$ are unitless as well.

\begin{figure}
    \centering
    \includegraphics[width=0.45\textwidth]{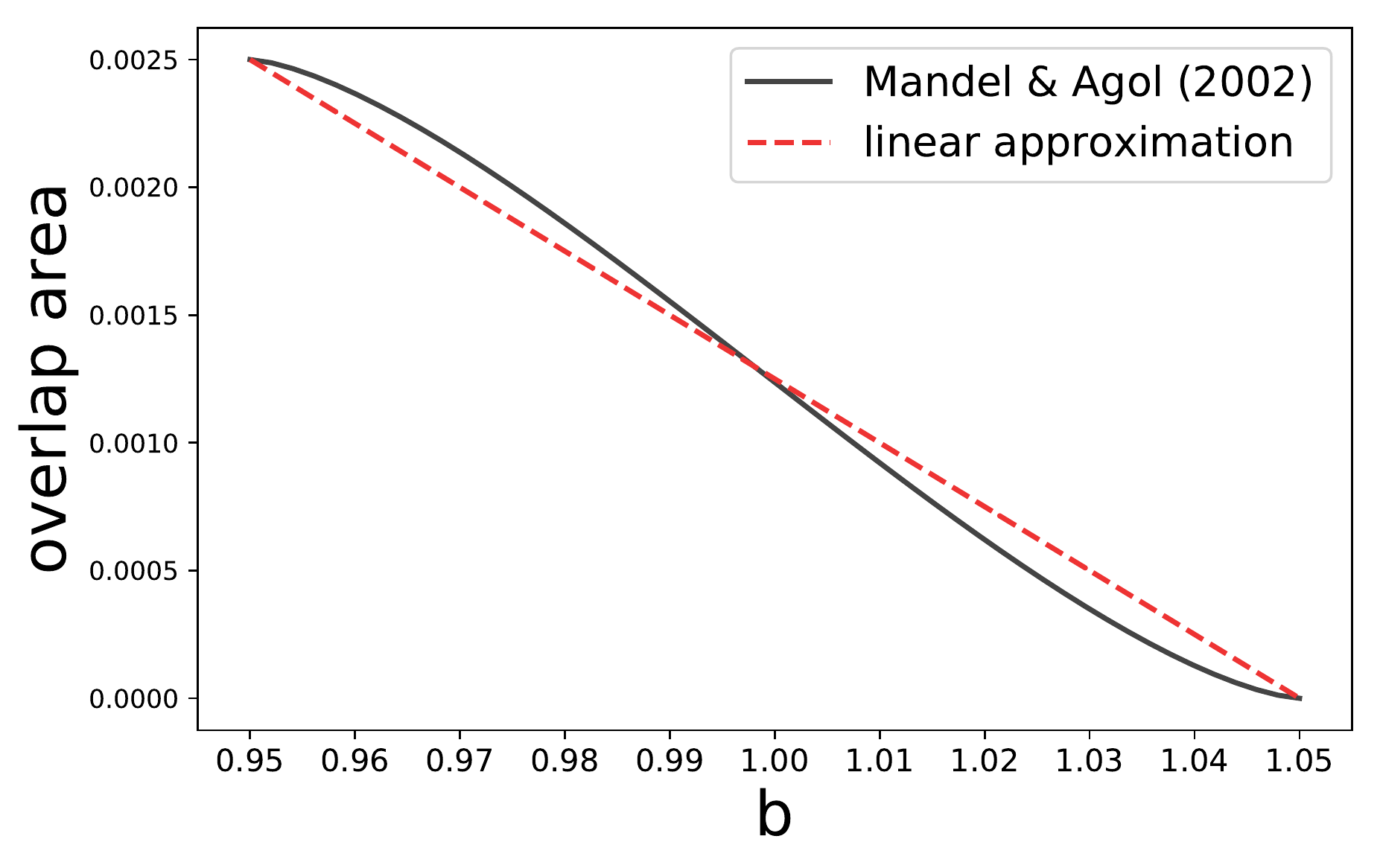}
    \caption{Planet-star overlap area at mid-transit as a function of impact parameter, $b$, for planet-to-star radius ratio $r = 0.05$. The plot is restricted to show only the grazing regime, i.e. $1-r < b < 1+r$. The solid black line shows the exact geometric solution derived by \citet{MandelAgol2002} and presented in this paper as Equation \ref{eq:mandel_agol}. Even though the full geometry is quite complicated, the resultant curve is well approximated by a simple linear function (dashed red line).}
    \label{fig:mandel_agol}
\end{figure}

The first quantity, $\lambda$, is derived from a linear approximation to the area of partial overlap between two spheres \citep{MandelAgol2002}; see Appendix \ref{appx:A} for details. Thus, $\lambda$ is closely related to the transit depth in the grazing regime. But, note that because $\lambda$ ranges over (0, $2r^2$) for grazing transits, the relation is closer to $\lambda \approx 2\delta$ than to $\lambda \approx \delta$. We caution the reader \underline{not} to use $\lambda$ as a basis parameter outside of the grazing regime because it is explicitly tied to the geometry of grazing transits. Figure \ref{fig:mandel_agol} demonstrates that the exact \citet{MandelAgol2002} geometry is well matched by a simple linear function $\lambda(b)$ at fixed r as long as $r < 1$, which will virtually always be the case for exoplanets orbiting main sequence or giant branch stars. We have not rigorously checked how the validity of our assumptions break down when $r \geq 1$, and so the results in this paper will likely need to be adjusted if they are to be applied to substellar companions of brown dwarfs \citep{Jung2018} or white dwarfs \citep{Vanderburg2020}.

The second quantity, $\gamma$, indicates the extent to which a transit is grazing or non-grazing, with transition occurring at $\gamma = 1$. When $\gamma \geq 1$, the transit is non-grazing; when $-1 < \gamma < 1$, the transit is grazing; when $\gamma \leq -1$ the planet does not transit at all. We refer to $\gamma$ as the \textbf{\textit{grazing coordinate}}, and in \S\ref{sec:umbrella} we will see that it plays a special role in our umbrella sampling routine.

When converting from one basis to another, care must taken in order to avoid inadvertently introducing unwanted priors. For a thorough discussion of the implicit priors introduced by our reparameterization and for a recipe to establish sensible prior distributions for $\lambda$ and $\gamma$, see Appendix \ref{appx:B}. The important point is that in addition to mapping $\{r,b\} \rightarrow \{\lambda,\gamma\}$, we add additional terms to the log-likelihood function as needed to ensure that our priors remain consistent between parameterizations.

By construction, our new $\{\lambda,\gamma\}$ basis is far more orthogonal than the fiducial $\{r,b\}$ basis is for grazing transits. Conversely, $\{\lambda,\gamma\}$ is far \textit{less} orthogonal than $\{r,b\}$ is for non-grazing transits. To achieve good sampler performance, we must therefore make the restriction $\gamma < 1$ when using our new basis and the restriction $\gamma \geq 1$ when using the old basis. We stress that our new $\{\lambda, \gamma\}$ parameterization is specifically designed with grazing transits in mind and should not be applied to non-grazing geometries.

For many transits, some fraction of the posterior distribution will be consistent with both grazing and non-grazing trajectories, so fitting a transit will require at least two independent sampling runs, one to sample the grazing regime using $\{\lambda,\gamma\}$ and the other to sample the non-grazing regime using $\{r,b\}$. Recombining independent posterior chains into a single posterior distribution can be performed using the statistical technique of umbrella sampling, which will be introduced in \S\ref{sec:umbrella}. For now, we will restrict our analysis to consideration of the grazing regime in order to compare the relative performance of the two basis sets.

\subsection{Performance of the $\{r,b\}$ vs $\{\lambda,\gamma\}$ basis}\label{subsec:basis_comparison}

The fiducial basis set we have used thusfar is $\{P, t_0, \ln r, b, \ln T, q_1, q_2\}$, which we now compare to our new parameterization $\{P, t_0, \ln\lambda, \gamma, \ln T, q_1, q_2\}$. As a shorthand, we will continue to refer to these as the $\{r,b\}$ and $\{\lambda, \gamma\}$ bases, respectively, although any actual sampling will always be performed using $\ln r$ and $\ln \lambda$ in place of $r$ or $\lambda$.

To compare the two basis sets, we simulate a low signal-to-noise transit of a warm-Jupiter orbiting a Sun-like star and sample from the posteriors using Hamiltonian Monte Carlo. Simulated photometry is shown in Figure \ref{fig:simulated_photometry} and ground truth parameter values are presented in Table \ref{tab:sim_parameters}. Our model setup and sampling routine both follow the procedure described in \S\ref{subsec:simulation_description}, modified to restrict samples to grazing geometries. For each basis set, we perform 100 independent MCMC runs using two chains run for 5000 tuning steps and 1000 draws, generating 2000 samples per run. Rather than merely setting a hard boundary at the grazing transition, we added a biasing potential, $\psi$, to the likelihood such that

\begin{equation}\label{eq:psi_bias}
    \psi(\gamma) =
    \begin{cases}
			1 + \gamma & \text{$\gamma \leq 0$}\\
            1 - \gamma & \text{$0 < \gamma < 1$}
    \end{cases}
\end{equation}

which has the effect of preferentially biasing posterior samples toward the middle of the grazing regime. The term is related to umbrella sampling, and the motivation behind its inclusion will become apparent in \S\ref{sec:umbrella}.

\begin{figure*}
    \centering
    \includegraphics[width=0.90\textwidth]{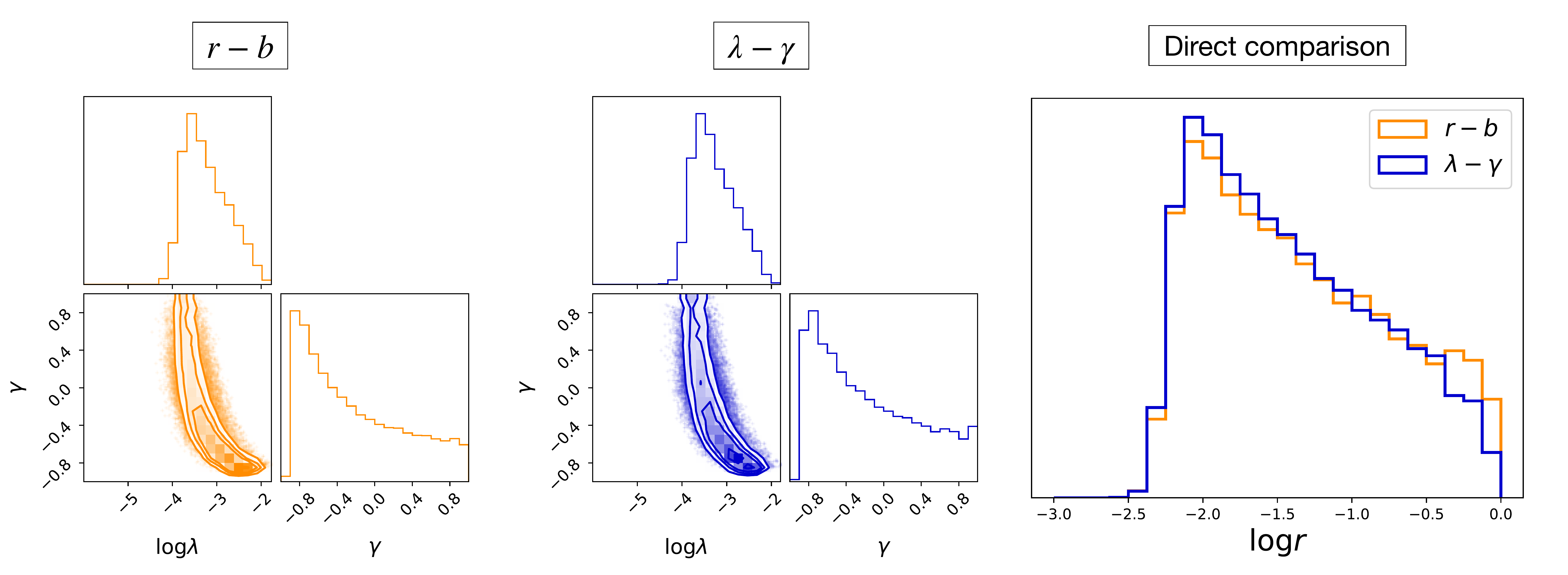}
    \caption{Posterior distributions of $\ln \lambda$, $\gamma$, and $\ln r$ for transit geometries restricted to the grazing regime for a near-grazing transit of a warm Jupiter orbiting a Sun-like star (model J-85). The effects of the biasing potential have been removed. See Table \ref{tab:sim_parameters} for ground truth simulation parameters and Figure \ref{fig:simulated_photometry} for the simulated photometry. The two parameterizations produce comparable posterior distributions, although the new $\lambda - \gamma$ basis performs $\sim20\%$ more efficiently. Note that for the left panel, $\lambda$ and $\gamma$ were not basis parameters, but were computed after the fact from samples of $r$ and $b$. A detailed discussion of the model parameterization is presented in \S\ref{sec:new_basis}.}
    \label{fig:basis_corner_compare}
\end{figure*}

Our new $\{\lambda,\gamma\}$ basis performs more efficiently than the standard $\{r,b\}$ basis and produces consistent posterior distributions. For a simulated near-grazing transit (model J-85), the total runtime for a given run using $\{r,b\}$ was $389 \pm 34$ seconds, compared to  $386 \pm 26$ seconds using $\{\lambda,\gamma\}$, a nearly identical wall clock time. On the balance, the autocorrelation length of the chains was a little longer when using the new basis compared to the standard basis, resulting in a larger number of effective samples obtained using our $\{\lambda,\gamma\}$ basis compared to the standard $\{r,b\}$ basis. Evaluated using the autocorrelation length for $r$, the time per effect sample was 1.6 seconds using $\{\lambda,\gamma\}$ vs 2.3 seconds using $\{r,b\}$, a $29\%$ gain in efficiency. We repeated this autocorrelation analysis using posterior chains for $b$ and $T$ finding a gains in efficiency of $28\%$ and $4\%$, respectively. when using our new basis. The relative performances of the two bases was comparable for various other simulated transit geometries (see Table \ref{tab:sim_parameters}), typically producing a $\sim20\%$ gain in efficiency for generating effective samples of $r$ and $b$ and a roughly equivalent efficiencies for generating effective samples of $T$. The two bases produce consistent posterior distributions (Figure \ref{fig:basis_corner_compare}). We conclude that our new $\{\lambda,\gamma\}$ basis will be preferred under most circumstances.

\section{Umbrella sampling}\label{sec:umbrella}

Umbrella sampling \citep{TorrieValleau1977} is a statistical tool designed for estimating complicated target distributions - e.g. multimodal distributions or degeneracy ridges - for which standard sampling techniques fail. Umbrella sampling does not replace existing sampling methods, but rather works in tandem with these methods to produce more robust posterior estimates. The basic idea is to split a complicated sampling problem into multiple smaller, more manageable problems, each restricted to a narrow region (or window, in the standard nomenclature) of parameter space. Samples are obtained separately from each window using whatever sampling technique the user prefers  - e.g. Hamlitonian Monte Carlo \citep{Neal2011, pymc3:2016} or ensemble sampling \citep{GoodmanWeare2010, ForemanMackey2013} - after which the samples are recombined into a single joint posterior distribution.

\begin{figure*}
    \centering
    \includegraphics[width=0.9\textwidth]{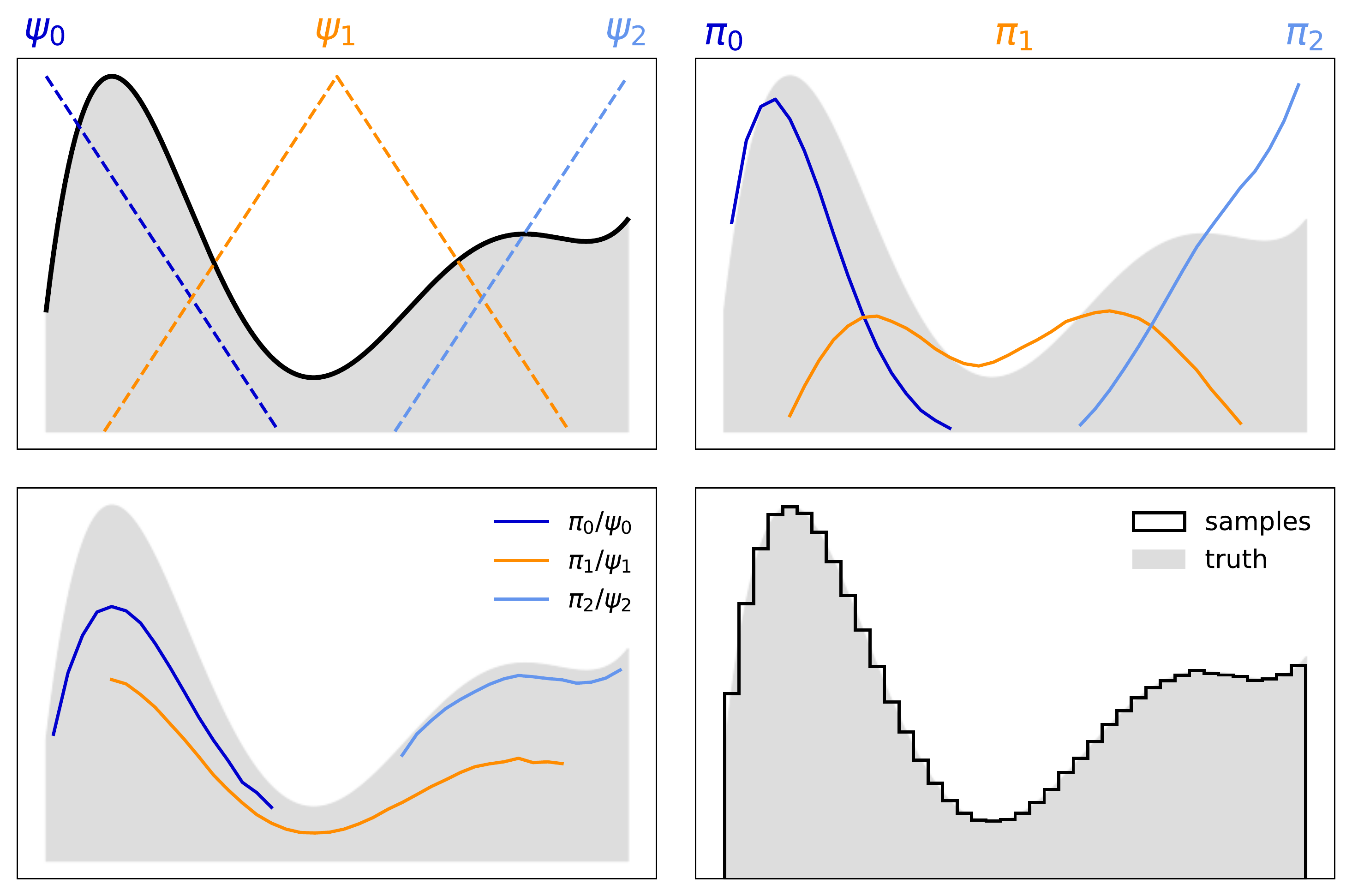}
    \caption{Schematic illustration of the umbrella sampling method, inspired by \citet{Smith2017}. The method is designed to facilitate sampling from multi-modal target distributions (shaded grey region in all panels). \textit{Top left}: The target distribution (emphasized with a thick black line) has low-probability valley which will create a bottleneck for standard sampling techniques. In order to ensure sampling from the full posterior space, we split the problem into three windows, each assigned a bias function, $\psi_i$. \textit{Top right}: After sampling independently from each window, we have three biased sub-distributions $\pi_i$. \textit{Bottom left}: Removing the effect of the bias functions, $\psi_i$ produces three unbiased sub-distributions with unknown offsets between one another. \textit{Bottom right}: calculating the window weights, $z_i$, and recombining all sub-distributions $\pi_i$ into a single joint posterior recovers the true target distribution. See \S\ref{subsec:umbrella} for a detailed discussion. A tutorial for reproducing this plot by implementing umbrella sampling can be found at \url{https://gjgilbert.github.io/tutorials/umbrella_sampling/}.}
    \label{fig:umbrella_tutorial}
\end{figure*}

Although umbrella sampling has rarely been applied to astrophysical problems, the technique is widely used in the field of molecular dynamics where it originated \citep{TorrieValleau1977}. The literature on umbrella sampling is extensive, but because most examples of its use are presented in the context of highly technical chemical analyses, there exists a precipitous barrier to entry for many astronomers (this paper's author included) who lack the domain expertise to easily comprehend the specialized scientific content surrounding the general statistical tool we wish to adopt. One goal of this paper is therefore to present an accessible, high-level introduction to umbrella sampling tailored toward the needs of astronomers in order to establish a gateway into the wider umbrella sampling literature. For a more rigorous introduction, we direct the interested reader to a recent review of umbrella sampling by \citet{Kastner2011}, as well as to the first astrophysical application of by \citet{Matthews2018}. Indeed, much of the pedagogy in this section was borrowed from \citet{Matthews2018} - particularly their \S 2.1 - and any astronomer wishing to implement umbrella sampling themselves will likely benefit from reviewing that paper in tandem with the present manuscript. Because learning to use new mathematical tools is often best accomplished through a ``hands-on'' approach, we have also developed a Python tutorial for implementing umbrella sampling, available at \url{https://gjgilbert.github.io/tutorials/umbrella_sampling/}.

\subsection{A brief overview of umbrella sampling}\label{subsec:umbrella}

Let us begin by assuming that we wish to sample from some arbitrary target distribution which possesses a complicated geometry (Figure \ref{fig:umbrella_tutorial}). Standard sampling techniques will do a poor job at traversing the low probability ``valleys'' between high probability ``peaks,'' resulting in poorly mixed posterior chains and incomplete sampling of the target distribution. One way around this issue is to add an additional bias term to the likelihood in order to ``level out'' the peaks and valleys, thereby simplifying the geometry. If the target distribution were known \textit{a priori} (which of course it is not), we could add a single bias term to the entire distribution to make it flat throughout. In practice, however, the more viable approach is to break the complicated target distribution into several overlapping \textit{\textbf{windows}}, sample separately from each window, and then recombine the sub-samples into a joint posterior distribution. Each window will be assigned its own \textbf{\textit{bias function}}, sometimes called simply a \textit{\textbf{bias}} or \textit{\textbf{umbrella}}. The bias functions serve to restrict the sampler to a given window and ensure that a significant fraction of samples are drawn from the low probability valleys.

Before defining our windows and biases, we must first identify a suitable variable, $x$, which we will use to construct a sampling framework. In the molecular dynamics literature, $x$ is usually called the \textit{reaction coordinate} because it corresponds to a real physical quantity related to chemical reactions such as free energy or molecular bond strength; in this manuscript we will refer to $x$ as the \textit{\textbf{umbrella coordinate}} (hence our earlier terminology for the grazing coordinate). The optimal choice of $x$ will be dictated by the geometry of the target distribution. For example, if the target distribution consists of several isolated peaks, $x$ could be defined along the line connecting those peaks. Such detailed advance knowledge is not strictly necessary, however, and in many cases it is possible to select a good (though perhaps sub-optimal) umbrella coordinate even for a blind search. In any case, the prior information needed to identify a suitable umbrella coordinate is comparable to the prior information needed to properly specify the model in the first place, and the choice of umbrella coordinate should follow from the structure of the problem. For a more in-depth discussion of strategies for choosing umbrella coordinates, particularly under information-limited circumstances, see \citet{Matthews2018}.

Once we have selected our umbrella coordinate, $x$, our next task is to define our window bounds and a set of $N$ corresponding bias functions, $\psi_i(x)$ (Figure \ref{fig:umbrella_tutorial}, top left panel). Once again, the optimal choice of windows and biases depends on the geometry, so the more that can be learned via exploratory analysis, the better. Fortunately, however, the results of umbrella sampling are insensitive to the particular choice of window bounds and bias functions provided that two conditions are met: (1) each window is adequately sampled, and (2) there is sufficient overlap between windows in order to allow accurate determination of relative window weights. We are thus free to define $\psi_i$ in whatever manner is most convenient for the problem at hand.

With windows and biases defined, we now sample from the target distribution, $\pi(x)$ separately from each of the $N$ windows, thereby producing $N$ biased posterior sub-distributions $\pi_i(x)$ (Figure \ref{fig:umbrella_tutorial}, top right panel). The sub-distributions relate to the (known) bias functions and to the (unknown) target distribution, via the equation

\begin{equation}\label{eq:pi_i}
    \pi_i(x) = \frac{1}{z_i}\psi_i(x)\pi(x)
\end{equation}

where $z_i$ are the window weights quantifying the relative contribution of each $\pi_i$ to the combined target distribution, $\pi$. Because each $\pi_i$ is a probability distribution, $\int\pi_i(x)dx=1$, and the window weights $z_i$ can be calculated via integration of Equation \ref{eq:pi_i} as

\begin{equation}\label{eq:zi_int}
    z_i = \int \psi_i(x)\pi(x)dx  = \langle \psi_i \rangle_{\pi}
\end{equation}

where $\langle \psi_i \rangle_{\pi}$ denotes the average of some function $f$ with respect to $\pi$. In other words, to determine $z_i$, we take the average of each $\psi_i$ weighted by the empirically sampled target distribution, $\pi$. If the full target distribution $\pi$ were known, calculating the window weights $z_i$ would be trivial. But of course $\pi$ is not known - it is precisely the quantity we are trying to determine! Furthermore, we don't actually have samples of $\pi$ yet. Rather, we have $N$ sets of biased sub-samples, $\pi_i$, meaning we will need to compute $\langle \psi_i \rangle_{\pi_j}$ for each $(i,j)$ and then combine these to estimate $\langle \psi_i \rangle_{\pi}$. The challenge is that this final combination step depends on $z$, making the whole process a bit circular. Once the $z_i$ are known, however, the biased sub-distributions, $\pi_i$, can be easily combined into a single joint posterior distribution, $\pi$ (Figure \ref{fig:umbrella_tutorial}, bottom panels).

Different methods for implementing umbrella sampling more or less come down to different strategies for solving the integral in Equation \ref{eq:zi_int}. The most popular method is the Weighted Histogram Analysis Method \citep[WHAM;][]{Kumar1992}, which works by binning the data and computing a histogram in the overlap region. Another popular method is the Multistate Bennet Acceptance Ratio \citep[MBAR;][]{ShirtsChodera2008}, which does not require discretization of the data. Both WHAM and MBAR can be derived from maximum likelihood or minimum asymptotic variance principles (see the references above for proofs). Recently, \citet{Thiede2016} and \citet{Dinner2017} demonstrated that the determination of umbrella weights $z_i$ can be recast as an eigenvector problem, a method which they term the Eigenvector Method for Umbrella Sampling (EMUS). Establishing umbrella sampling as an eigenvector problem has the twin advantages of being computationally efficient and facilitating accurate error analysis, and so we adopt EMUS as our method of choice here.

Following \citet{Matthews2018}, we restate Equation \ref{eq:zi_int} as an explicit sum

\begin{equation}\label{eq:zi_sum}
    z_j = \sum_{i=1}^N \Bigg\langle \frac{\psi_j(x)}{\sum_{k=1}^N \psi_k(x)/z_k} \Bigg\rangle_{\pi_i}
\end{equation}

where $\langle \rangle_{\pi_i}$ denotes an average with respect to $\pi_i$. Because the umbrella weights $z_i$ enter the equation both on the left-hand side of the equation and in the denominator sum on the right-hand side, Equation \ref{eq:zi_sum} must be solved iteratively. To do so using EMUS, we first define a square \textit{overlap matrix}, $F$, with each element $(i,j)$ defined as

\begin{equation}\label{eq:Fij}
    F_{ij} = \Bigg\langle \frac{\psi_j/z_i}{\sum_{k=1}^N \psi_k/z_k} \Bigg\rangle_{\pi_i}
\end{equation}

As its name implies, $F$ tracks the extent to which samples drawn within one window fall under the umbrella of any other window. On diagonal terms will usually have larger values (because all samples drawn from window $i$ by construction fall under umbrella $\psi_i$), and when windows $(i,j)$ do not overlap, $F_{ij} = F_{ji} = 0$.

In order to calculate $z_i$ using linear algebra, we first define $z \equiv [z_1,z_2,...,z_N]$ as a vector. Taking the product of $z$ and the $j^{\rm th}$ column of $F$ yields

\begin{equation}\label{eq:emus_multiplication}
\sum_{i=1}^N z_i F_{ij} = \sum_{i=1}^N \Bigg\langle \frac{\psi_j}{\sum_{k=1}^N \psi_k/z_k} \Bigg\rangle_{\pi_i} = \langle \psi_j \rangle_{\pi}
\end{equation}

Recall from Equation \ref{eq:zi_int} that $z_j = \langle \psi_j \rangle_{\pi}$, so $\sum_i z_i F_{ij} = z_j$. Considering all columns in $F$ simultaneously yields the left eigenvalue problem 

\begin{equation}\label{eq:emus_eigenvector}
zF = z
\end{equation}

which when solved provides an estimate of the window weights.

If we knew $F$ \textit{a priori}, finding the eigenvalues and eigenvectors of Equation \ref{eq:emus_eigenvector} would be a straightforward application of linear algebra. But, in practice, we need to estimate both $z$ and $F$ from our empirical samples, $\pi_i$. As we noted earlier, this must be done iteratively. Our strategy will be to pick a starting guess for $z$ and calculate a first estimate of $F$ from Equation \ref{eq:Fij}. We'll then use our estimate of $F$ to calculate an updated value for $z$ using Equation \ref{eq:emus_eigenvector}, and then iterate between Equations \ref{eq:Fij} and \ref{eq:emus_eigenvector} until the result converges. In practice the problem is often nearly converged after just one or two iterations, and both the final result and convergence rate are insensitive to the particular starting estimate of $z$.

In summary, the steps of umbrella sampling are: (1) choose a suitable umbrella coordinate $x$, (2) define windows and biases $\psi_i$, (3) sample from each window to produce biased sub-distributions $\pi_i$, (4) calculate window weights $z_i$ by iteratively solving Equations \ref{eq:Fij} and \ref{eq:emus_eigenvector}, and finally (5) recombine all sub-samples into the joint posterior estimate $\pi$ by inverting Equation \ref{eq:pi_i}. Note that unlike standard direct sampling methods which produce a single set a unweighted samples, umbrella sampling produces multiple sets of weighted samples (with weights given by $z_i$), and these weights must be taken into account when estimating posterior distributions or summary statistics.

A Python tutorial for implementing EMUS can be found at \url{https://gjgilbert.github.io/tutorials/umbrella_sampling/}.

\subsection{Applying umbrella sampling to the transit model}\label{subsec:full_model}

We now introduce our full umbrella sampling routine as applied to the transit fitting problem. Properly implemented, our new method produces posterior estimates which are more accurate than estimates obtained using standard direct sampling techniques. The key components of our method are (1) splitting the transit fitting problem into separate windows for grazing vs non-grazing geometries and (2) adopting a unique parameter basis within each window tailored to the specific geometry at hand.

For our umbrella coordinate we adopt the grazing coordinate, $\gamma \equiv (1-b)/r$, which was introduced in \S\ref{subsec:basis_specification}. Defining our windows in terms of $\gamma$ allows us to easily separate posterior sampling into grazing and non-grazing runs, with the cutoff occurring at $\gamma = 1$. While developing our method, we first attempted to implement a simple two-umbrella scheme wherein the non-grazing window extended slightly into the grazing regime and, conversely, the grazing window extended slightly into non-grazing regime. However, we found that the sampler still often became stuck at the grazing to non-grazing transition, leading to poorly mixed chains and inaccurate results. We therefore found it necessary to restrict the grazing umbrella to strictly grazing geometries ($\gamma < 1)$ and the non-grazing umbrella to strictly non-grazing geometries ($\gamma > 1$). Windows must have at least some overlap with their neighbors, so we introduced a third ``transition'' umbrella centered on the grazing to non-grazing boundary at $\gamma = 1$ and extending a little way into both the grazing and non-grazing regimes in order to bridge the gap. We find that this simple three-umbrella scheme performed well under a wide range of circumstances.

\begin{figure}
    \centering
    \includegraphics[width=0.45\textwidth]{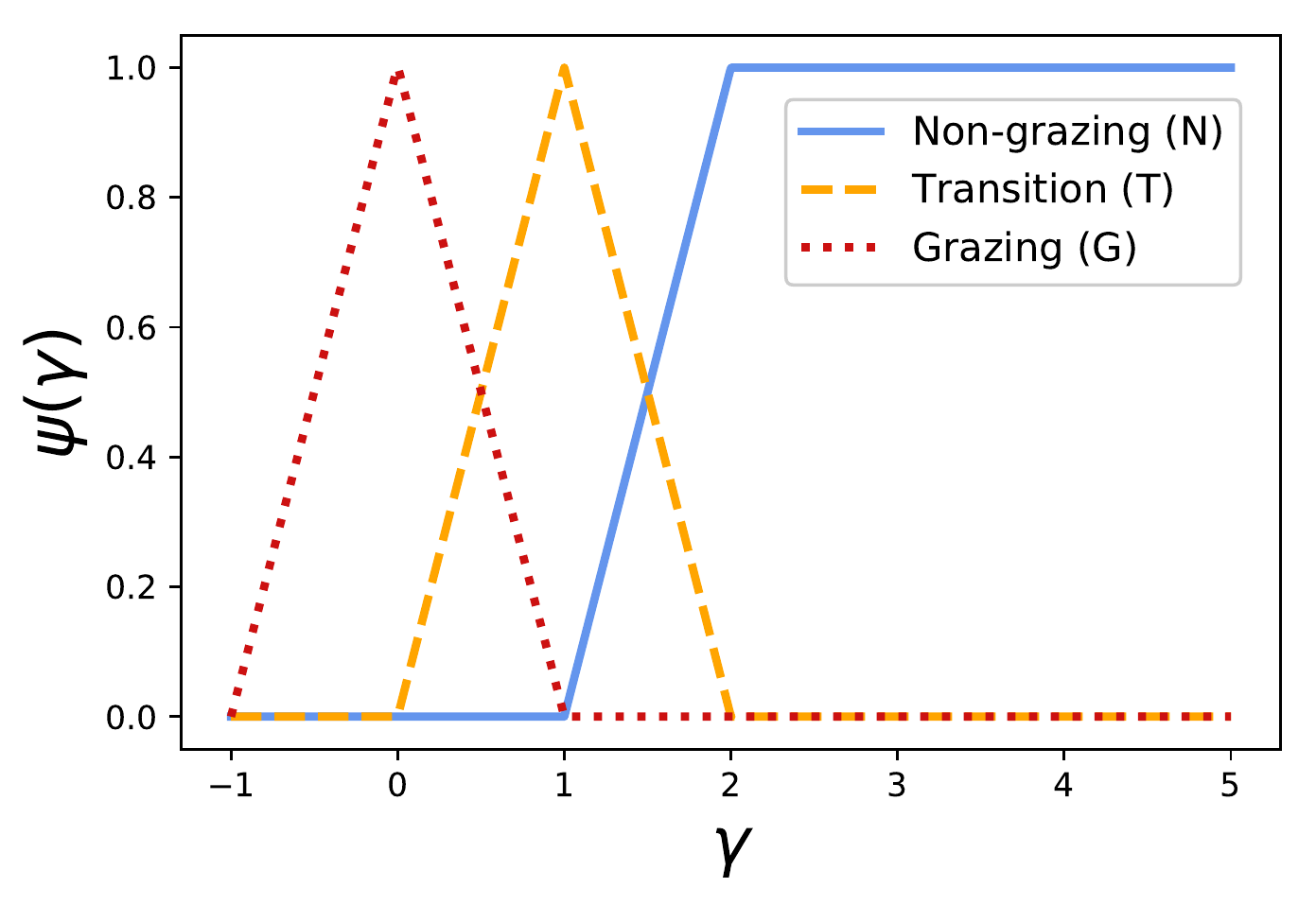}
    \caption{Our umbrella bias functions, $\psi(\gamma)$. The solid blue line is the non-grazing umbrella ($\psi_N$). The dashed orange line is the transition umbrella ($\psi_T$). The dotted red line is the grazing umbrella  ($\psi_G$). We have opted to use tent biases because these are simple to perform calculations with, but because umbrella sampling is insensitive to the particular choice of bias functions \textemdash provided that windows overlap \textemdash many other functional forms would perform just as well.}
    \label{fig:umbrella_functions}
\end{figure}

We define our bias functions over the non-grazing (N), transition (T), and grazing (G) windows as

\begin{equation}\label{eq:psi_N}
    \psi_N \simeq 
    \begin{cases}
            \gamma - 1 & \text{$1 < \gamma < 2$} \\
			1 & \text{$\gamma \geq 2$}
    \end{cases}
\end{equation}

\begin{equation}\label{eq:psi_T}
    \psi_T \simeq 
    \begin{cases}
			\gamma & \text{$0 \leq \gamma < 1$}\\
            2 - \gamma & \text{$1 \leq \gamma < 2$}
    \end{cases}
\end{equation}

\begin{equation}\label{eq:psi_G}
    \psi_G \simeq 
    \begin{cases}
			1 + \gamma & \text{$\gamma \leq 0$}\\
            1 - \gamma & \text{$0 < \gamma < 1$}
    \end{cases}
\end{equation}

where the symbol ``$\simeq$'' denotes that normalization constants have been omitted. These biases are shown graphically in Figure \ref{fig:umbrella_functions}. We have opted to use tent biases out of mathematical convenience, but as noted above, the results of umbrella sampling are in general insensitive to any particular choice of bias function. The reader is thus free to chose any other bias function if they so desire. However, we do caution that while the shape of the bias within each window is mostly unimportant, altering the window widths (i.e. the range of $\gamma$ spanned by each $\psi$) can have a significant effect. Indeed, while developing this method we undertook considerable effort to ensure that windows overlapped enough to facilitate calculation of the window weights without being so wide as to lead to geometric degeneracies. We therefore advise that anyone attempting to apply our method should only adjust the window bounds after careful consideration of the consequences. Unless one has a strongly motivated reason to alter the windows, the safest approach is to stick with the limits presented in Equations \ref{eq:psi_N} - \ref{eq:psi_G}.

Because each window will be sampled independently of the others, we are free to use a different parameter basis within each window. Motivated by the results of \S\ref{subsec:basis_comparison}, we adopt the standard $\{\ln r, b\}$ basis for the non-grazing umbrella, while for the grazing umbrella we adopt our new $\{\ln\lambda, \gamma\}$ basis. For sampling runs under the transition umbrella, we adopted the hybrid parameter basis $\{\ln r, \gamma\}$, which we found worked well over a range of conditions. Each basis set is tailored to the specific geometry within its corresponding window, and thus performs well under its own umbrella. Umbrellas are always defined in terms of $\gamma$, but $\gamma$ is not always a basis parameter. In practice, in order to define our window bounds and bias functions, we must first calculate $\gamma$ from any two out of three basis parameters $\{r, b, \lambda\}$ following Equation \ref{eq:lam_gam}.

The stage is now set, and at this point one could in principle draw samples from each window and then recombine them into a final joint posterior following the weighting prescription described in \S\ref{subsec:umbrella}. However, there is one final complication that must be addressed first, namely that we do not know the transit geometry ahead of time and so we cannot be sure whether samples $\pi_T$ obtained under the T umbrella will overlap with both $\psi_G$ and $\psi_N$. Recall that umbrella sampling does not merely require that windows overlap, but instead imposes the more stringent requirement that at least some samples obtained within each window fall into the overlap region with their neighboring windows. This subtle condition demands careful attention, but turns out to be a blessing in disguise.

In order to ensure that our window weights will be properly determined, we first sample from the transition umbrella, producing a (biased) posterior distribution $\pi_T(\gamma)$ with samples restricted by $\psi_T$ to lie between $0 < \gamma < 2$. Because $\gamma$ tells us how strongly grazing the transit is and because all samples $\pi_T(\gamma)$ must by construction fall near the grazing/non-grazing transition boundary, we can use $\pi_T(\gamma)$ to infer whether the transit geometry is grazing or not. If all samples $\pi_T(\gamma)$ have $\gamma < 1$, we can be confident that the planet is on a grazing trajectory; conversely if all samples $\pi_T(\gamma)$ have $\gamma > 1$, we can be confident that the planet is on a non-grazing trajectory. In the former case (all $\gamma < 1$) we then need only draw samples from the grazing window, whereas in the latter case (all $\gamma > 1$) we need only draw samples from the non-grazing window. In fact, under these circumstances umbrella sampling may no longer be needed, as the N or G windows will by themselves cover the full span of the relevant parameter space. However, at this point samples from the T umbrella have already been obtained, so one may as well proceed with a two-umbrella scheme.

We recommend that all future transit modeling efforts \textemdash\ even those which do not intend use umbrella sampling for their final analysis \textemdash\ first conduct an exploration of the grazing/non-grazing transition boundary, aided by $\psi_T$ to ensure adequate sampling of the region immediately surrounding $\gamma = 1$. Depending on the circumstances, one may wish to set a more or less lenient condition for categorizing a transit as grazing/non-grazing than we have proposed here (i.e. all $\gamma < 1$ vs all $\gamma > 1$), but the core strategy would remain the same. Conclusively ruling in/out grazing geometries will afford us greater confidence in results derived from transiting modeling, and if widely adopted we anticipate our ``check the transition region first'' approach will reveal previously unnoticed inaccuracies or systematic offsets in transiting exoplanet catalogs.

Samples may be drawn using any suitable sampling method, and provided that all posterior chains are well mixed and pass the necessary convergence checks, the choice of sampler will be inconsequential to the final results, save perhaps a difference in computational efficiency. Once we have drawn samples from all three windows (or perhaps only two, if $\pi_T(\gamma)$ rules out one geometry or another), calculation of the window weights, $z_i$, is a straightforward application of the EMUS algorithm presented in \S\ref{subsec:umbrella}; once $z_i$ have been calculated, we can then immediately estimate the posterior distributions and summary statistics.

\section{Comparison of results from standard sampling techniques to umbrella sampling}\label{sec:sampler_comparison}

We now test our proposed method by simulating transit lightcurve photometry for several prototypical star-planet configurations and then comparing posterior inferences obtained via umbrella sampling to inferences obtained using a standard direct sampling approach. Throughout these tests, we follow the same data simulation procedure and Hamiltonian Monte Carlo sampling routine described in \S\ref{subsec:simulation_description}, modified to incorporate a moderately informative prior on eccentricity. Rather than incorporating $e$ and $w$ as free parameters in our model, we instead inferred these quantities using the photoeccentric effect \citep{FordQuinnVeras2008, DawsonJohnson2012}, thus necessitating priors on both $e$ and $\rho_{\star}$. For $e$ we assumed a Rayleigh distribution with scale parameter $\sigma_e = 0.21$, corresponding to the single-planet value found by \citet{Mills2019}; for $\rho_{\star}$, we assumed a $10\%$ Gaussian measurement uncertainty. In practice, placing priors on $e$ and $\rho$ serves to place indirect priors on $T$ and $b$. We will address the role of eccentricity priors and describe the effects of several alternative prior distributions in greater detail is \S\ref{sss:eccentricity} below. 

We perform three tests of our method, each focused on a different star-planet architecture. In the first test (the ``J'' models; see Table \ref{tab:sim_parameters}), we place a warm Jupiter in orbit around a Sun-like star at various impact parameters in order to simulate grazing, near-grazing, and non-grazing trajectories. In the second test (models ``SE'' and ``MN'') we place a super-Earth and mini-Neptune on 21 day orbits around a star typical of the Kepler field, with inclinations scaled to produce comparable transit depths. In the third (model ``MHZ''), we place a rocky planet in the habitable zone of an M dwarf. Simulated photometry is shown in Figures \ref{fig:simulated_photometry}, \ref{fig:simulated_photometry_valley}, and \ref{fig:simulated_photometry_mhz}, and ground truth parameter values for each simulated lightcurve are collected in Table \ref{tab:sim_parameters}. As before, the important point throughout is that we have endeavored to simulate unremarkable transits, which we then model using techniques which are intended to be as uncontroversial as possible.

\subsection{A giant planet orbiting a solar twin}\label{subsec:case_study_J}

For our first test, we placed a warm Jupiter ($r=0.103$) on a circular orbit around a Sun-like star at three different impact parameters in order to create a grazing ($b=1.00$, model J-100), near-grazing ($b=0.85$, model J-85), and non-grazing ($b=0.22$, model J-22) trajectory. The transit duration for all three cases was set to $T=3.0$ hrs and then the orbital period was calculated in order to preserve $e=0$, resulting in orbital periods of 44.9, 13.0, and 3.6 days, respectively. In order to produce a comparable signal-to-noise, the simulated Gaussian noise for the grazing transit (J-100) was reduced by a factor of two relative to the non-grazing and near-grazing transits. The simulated photometry for all three configurations is shown in Figure \ref{fig:simulated_photometry}.

\subsubsection{Simulation J-85: a near-grazing transit}

\begin{figure}
    \centering
    \includegraphics[width=0.45\textwidth]{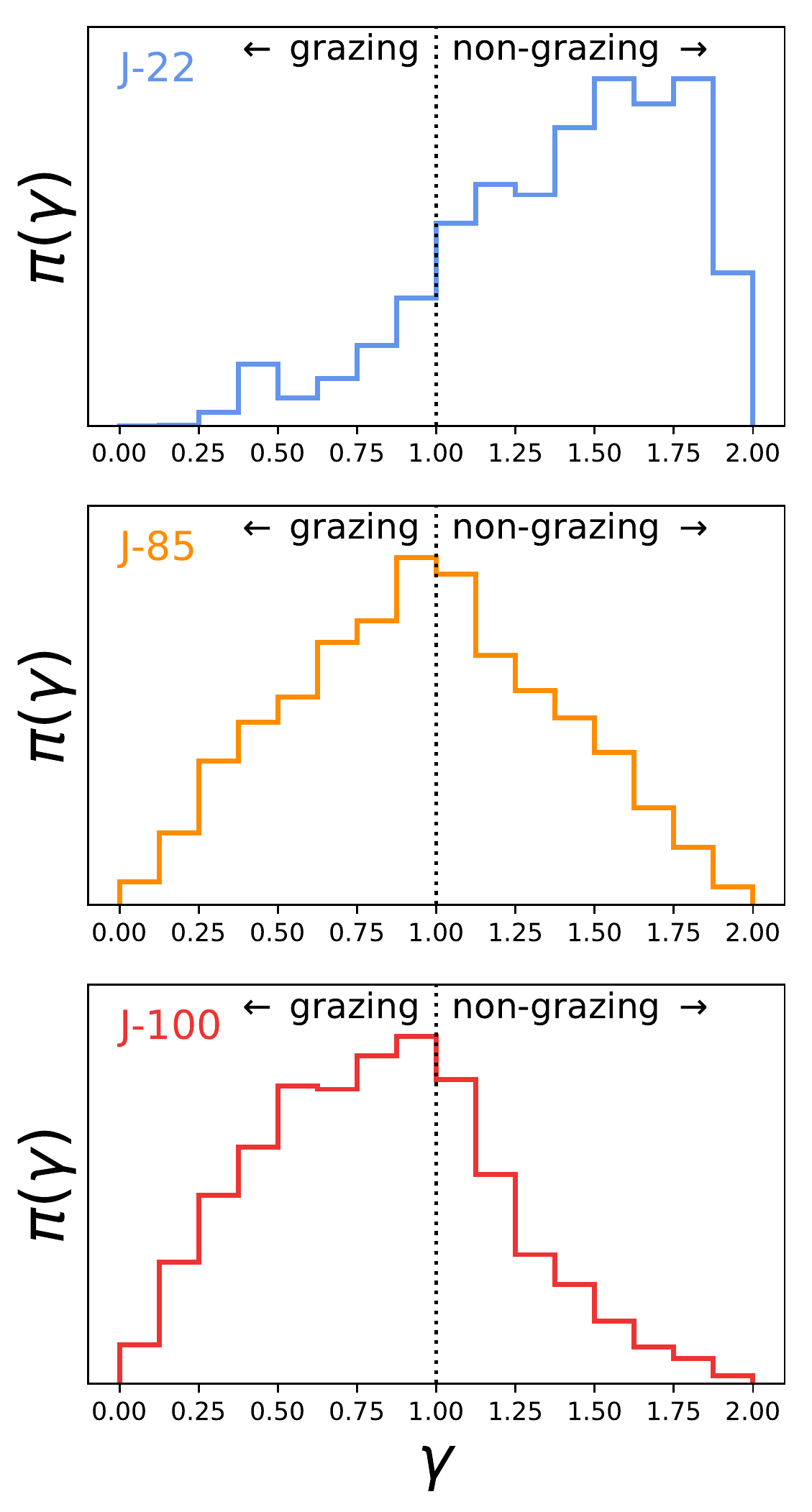}
    \caption{Distribution of the grazing coordinate, $\gamma$, for posterior MCMC samples obtained under the transition umbrella, $\psi_T$, for three simulated transits of a warm Jupiter orbiting a Sun-like star at various impact parameters. Simulated parameter values are collected in Table \ref{tab:sim_parameters} and corresponding simulated lightcurves are shown in Figure \ref{fig:simulated_photometry}. Unsurprisingly, the fraction of posterior samples consistent with a non-grazing geometry is highest for the simulated non-grazing transit (top), and vice-versa for a grazing geometry (bottom). The near-grazing transit (middle) reflects an intermediate state. In all three cases, at least some fraction of the posteriors are consistent with both a grazing and a non-grazing trajectory, indicating the transit geometry is ambiguous and the application of umbrella sampling is warranted.}
    \label{fig:transition_umbrella_J}
\end{figure}

We being by placing our warm Jupiter on a $P=13$ day orbit around its host star with $b=0.85$, thereby producing a transit chord that is non-grazing yet close enough to the stellar edge that limb darkening becomes significant. In this near-grazing regime, the transit morphology begins to shift from U-shaped to V-shaped, so we expect that some fraction of the posterior distribution will be consistent with both a grazing and non-grazing trajectory. Examination of $\gamma$ samples obtained under the transition umbrella, $\psi_T$, confirm that this is indeed the case (Figure \ref{fig:transition_umbrella_J}), validating our assertion that umbrella sampling is warranted.

\begin{figure*}
    \centering
    \includegraphics[width=0.9\textwidth]{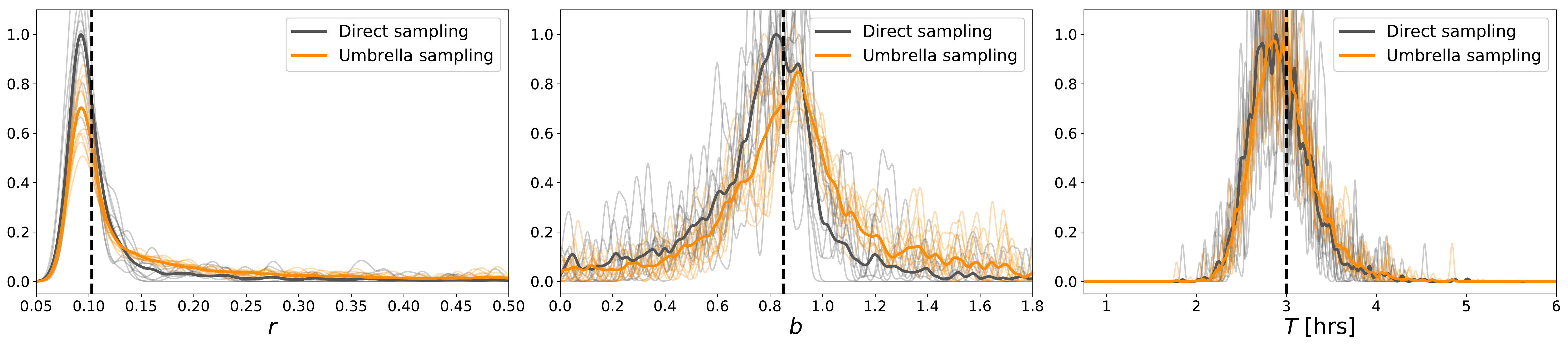}
    \caption{Posterior distributions of $r$, $b$, and $T$ for a simulated near-grazing transit of a Jupiter-size planet on a 13 day orbit around a Sun-like star (simulation J-85). See Table \ref{tab:sim_parameters} for simulated model parameters and Figure \ref{fig:simulated_photometry} for the simulated photometry. Each thin line represents a 2000 sample chain from a single independent Monte Carlo run, while the thick lines give the combined results of 20 such runs. Vertical dashed lines represent ground-truth parameter values. Both methods produce posterior distributions consistent with the true value, but only umbrella sampling is able to fully explore the high-$b$, high-$r$ tail of the distribution.}
    \label{fig:posteriors_J85}
\end{figure*}

Both direct sampling and umbrella sampling produce comparable distributions for $T$ and broadly similar estimates of $r$ and $b$ (Figure \ref{fig:posteriors_J85}). However, direct sampling does not fully explore the high-$b$, high-$r$ tail of the distribution. By-eye the differences appear slight, but the consequences of these skewed distributions become apparent when one calculates the marginalized $1\sigma$ uncertainties for $r$ and $b$. From direct sampling, we estimate $r = 0.098^{+0.042}_{-0.013}$, whereas from umbrella sampling, we estimate $r = 0.108^{+0.187}_{-0.021}$ (based on the $16^{\rm th}$, $50^{\rm th}$, and $84^{\rm th}$ percentiles). Although one might naively prefer the narrower posterior obtained via direct sampling, this misleadingly tight constraint on $r$ is predicated on the false assumption that the high-$b$ high-$r$ tail has been ruled out, when in fact it has simply not been explored. Umbrella sampling, on the other hand, ensures that the difficult to explore regions of the posterior have indeed been adequately sampled.

\subsubsection{Simulation J-22: a non-grazing transit}

\begin{figure*}
    \centering
    \includegraphics[width=0.9\textwidth]{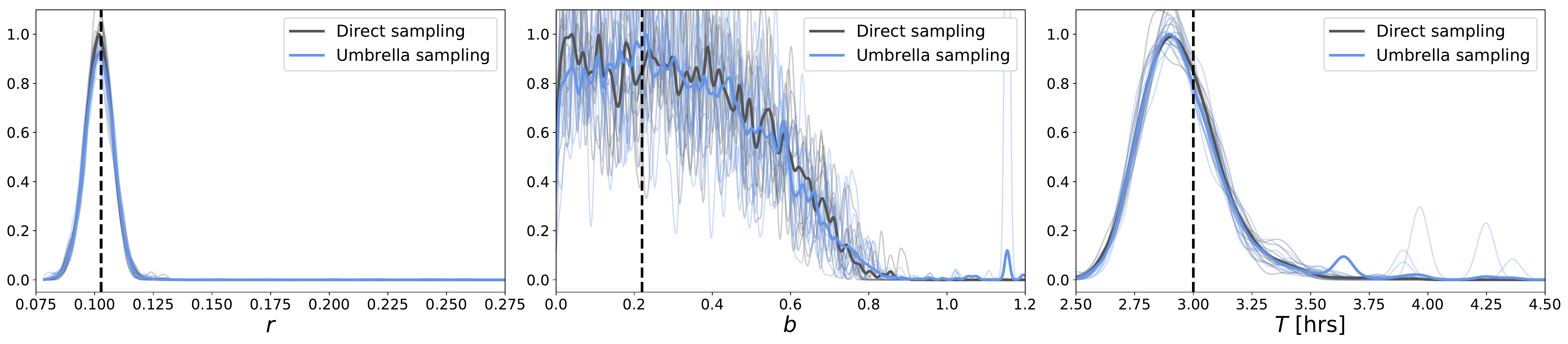}
    \caption{Posterior distributions of $r$, $b$, and $T$ for a simulated non-grazing transit of a Jupiter-size planet on a 3.6 day orbit around a Sun-like star (simulation J-22). See Table \ref{tab:sim_parameters} for simulated model parameters and Figure \ref{fig:simulated_photometry} for the simulated photometry. Each thin line represents a 2000 sample chain from a single independent Monte Carlo run, while the thick lines give the combined results of 100 such runs. Vertical dashed lines represent ground-truth parameter values. Both methods produce comparable results, however only umbrella sampling is able to smoothly explore deep into the grazing regime. This augmented exploration allows us to confidently rule out a grazing geometry by placing reliable upper limits on $r$ and $b$.}
    \label{fig:posteriors_J22}
\end{figure*}

We next modify our simulated transit by changing the impact parameter to $b=0.22$ in order to place the planet on a non-grazing trajectory. In order to keep the transit duration consistent at $T=3$ hrs, we shifted the orbital period to $P=3.6$ days. In this case, the results of the two methods are entirely consistent with one another (Figure \ref{fig:posteriors_J22}), as expected for a planet with negligible posterior mass consistent with a grazing geometry. Because there is a small but non-zero fraction fraction of samples with $b > 1-r$ (Figure \ref{fig:transition_umbrella_J}), trusting the results from direct sampling hinges on the implicit assumption that the sampler did not explore the grazing regime because the model and data are poorly matched there, rather than because the sampler encountered a bottleneck at the grazing/non-grazing boundary. The advantage of using umbrella sampling is that we can be more confident is our inferences because the sampler explores smoothly deep into the grazing regime, allowing us to be sure that the posterior likelihood there is indeed small.

\subsubsection{Simulation J-100: A grazing transit}

For our last test we shift the transit to $b=1.0$ in order to create a grazing trajectory (model J-100). Once again, we preserve the transit duration at $T=3$ hrs by adjusting the orbital period, in this case to to $P=45$ days. In order to compensate for the reduced transit depth of the grazing geometry, we reduce the photometric noise level by a factor of two, which gives this simulated transit (J-100)  a similar signal-to-noise ratio compared to the first two simulations (J-85 \& J-22). 

\begin{figure*}
    \centering
    \includegraphics[width=0.9\textwidth]{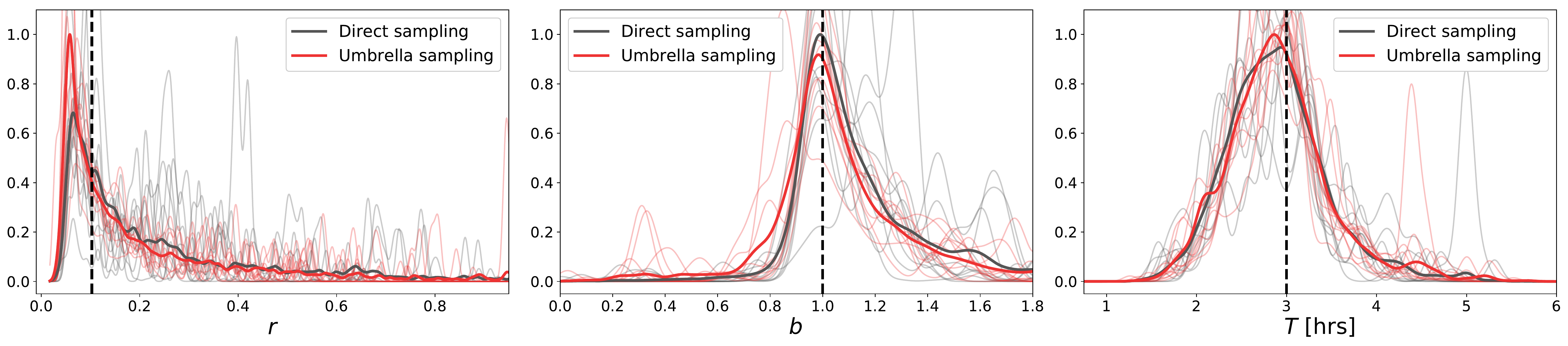}
    \caption{Posterior distributions of $r$, $b$, and $T$ for a simulated grazing transit of a Jupiter-size planet on a 45 day orbit around a Sun-like star (simulation J-100). See Table \ref{tab:sim_parameters} for simulated model parameters and Figure \ref{fig:simulated_photometry} for the simulated photometry. Each thin line represents a 2000 sample chain from a single independent Monte Carlo run, while the thick lines give the combined results of 100 such runs. Vertical dashed lines represent ground-truth parameter values. In this case, both mehods produce comparable results.}
    \label{fig:posteriors_J100}
\end{figure*}

The performance of the two methods for fitting a grazing transit is quite similar (Figure \ref{fig:posteriors_J100}). From direct sampling, we estimate $r = 0.148^{+0.262}_{-0.077}$,  $b = 1.05^{+0.28}_{-0.11}$; whereas from umbrella sampling, we estimate $r = 0.130^{+0.265}_{-0.070}$, $b = 1.03^{+0.29}_{-0.13}$. As with the non-grazing case (simulation J-22), the main advantage of umbrella sampling is that we can be sure we have explored the full posterior geometry, lending us greater confidence in our results.

\subsubsection{The role of eccentricity priors}\label{sss:eccentricity}

In order to investigate the effect of eccentricity priors, we repeated the experiment for the near-grazing transit (simulation J-85) using three additional eccentricity priors distributions. For the first two, we again used a Rayleigh prior, but now with scale parameter $\sigma_e = 0.0355$ or $\sigma_e = 0.008$. The former corresponds to the value found by \citet{Mills2019} for multiplanet systems, while the latter corresponds to the value by \citet{LithwickXieWu2012} for systems exhibiting large-amplitude transit timing variations. Recall that our original test used $\sigma_e = 0.21$, the \citet{Mills2019} single planet value. Our fourth and final test placed uniform (i.e. uninformative) priors on $e$. In all cases, we assumed a 10\% Gaussian measurement uncertainty on $\rho_{\star}$. Both methods show a similar sensitivity to choice of eccentricity prior (see Table \ref{tab:ecc_priors}).

\begin{table}[]
\renewcommand{\arraystretch}{2.0}
    \centering
    \begin{tabular}{l c c}
         Prior distribution & Direct Sampling & Umbrella Sampling \\
         \hline
         Rayleigh, $\sigma_e=0.008$  & $0.108^{+0.037}_{-0.014}$ & $0.118^{+0.198}_{-0.021}$ \\ 
         Rayleigh, $\sigma_e=0.0355$ & $0.108^{+0.038}_{-0.014}$ & $0.120^{+0.208}_{-0.022}$ \\ 
         Rayleigh, $\sigma_e=0.21$   & $0.098^{+0.042}_{-0.013}$ & $0.108^{+0.187}_{-0.021}$ \\ 
         Uniform,  $e \sim (0,1)$    & $0.091^{+0.033}_{-0.010}$ & $0.097^{+0.136}_{-0.015}$ \\ 
    \end{tabular}
    \caption{Marginalized MCMC posterior values for the planet-to-star radius ratio, $r$, of a simulated transit (simulation J-85), assuming four different eccentricity prior distributions. The true value is $r=0.103$. Posterior values quoted in this table correspond to the retrieved 16th, 50th, and 84th percentiles of $r$, with results arranged from most informative prior (top) to least informative (bottom). See text of \S\ref{sss:eccentricity} for discussion.}
    \label{tab:ecc_priors}
\end{table}

\subsection{A pair of planets straddling the radius valley}\label{subsec:case_study_radius_valley}

\begin{figure}
    \centering
    \includegraphics[width=0.45\textwidth]{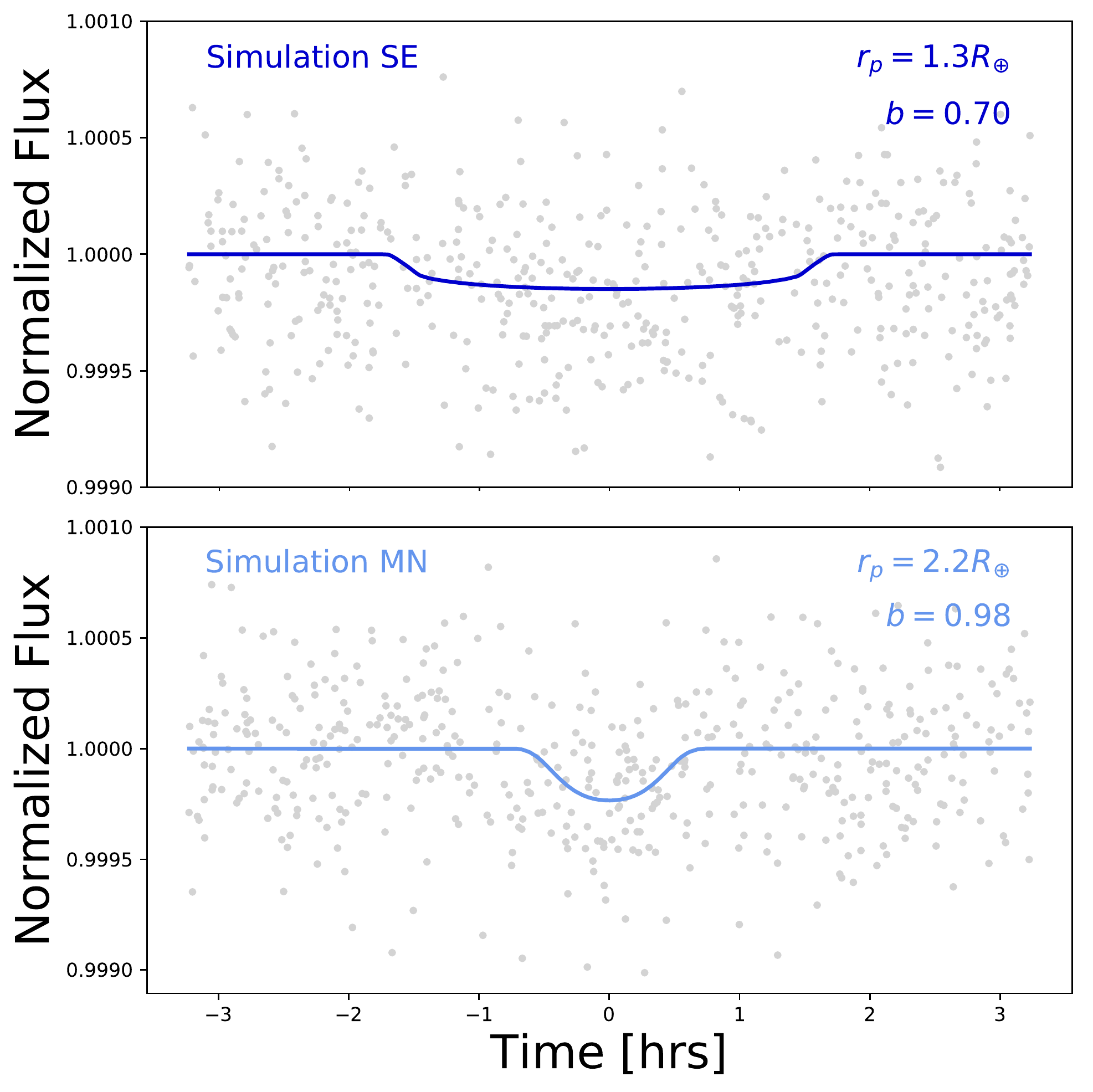}
    \caption{\textit{Top panel}: Simulated photometry for a mini-Neptune (model MN) orbiting a K-dwarf star on a barely grazing orbit. \textit{Bottom panel}: Simulated photometry for a super-Earth (model SE) orbiting the same star on a non-grazing orbit. Ground truth simulation parameters are collected in Table \ref{tab:sim_parameters}. See \S\ref{subsec:case_study_radius_valley} for discussion.}
    \label{fig:simulated_photometry_valley}
\end{figure}

A primary motivation for developing our umbrella sampling method is to accurately determine the radii of exoplanets in or near the radius valley \citep{Fulton2017}. More specifically, we would like to be able to measure the size of planets with $r_p \approx 1.6 R_{\oplus}$ and periods $P \lesssim 100$ days orbiting FGK stars, i.e. planets typical of the Kepler and \textit{K2} samples. For this case study, we simulate the transits of a pair of planets, each on a circular 21 day orbit around a K dwarf ($R_{\star}=0.92 R_{\odot})$. The first planet (simulation SE; a super-Earth) has $r_p = 1.3 R_{\oplus}$ and a non-grazing trajectory ($b=0.70)$. The second planet (simulation MN; a mini-Neptune) has $r_p = 2.2 R_{\oplus}$ and a barely grazing trajectory ($b=0.98)$. These setups produce a pair of transits with comparable transit depths, albeit distinct transit durations (Figure \ref{fig:simulated_photometry_valley}). For these test cases, an accurate estimate of $r$ thus hinges on accurate estimates of both $T$ and $b$. Our goal then is to investigate whether our competing sampling methods can constrain these three parameters with sufficient reliability to determine whether each planet exists on the rocky or gaseous edge of the radius valley.

\subsubsection{Simulation SE: A non-grazing super-Earth}

\begin{figure*}
    \centering
    \includegraphics[width=0.9\textwidth]{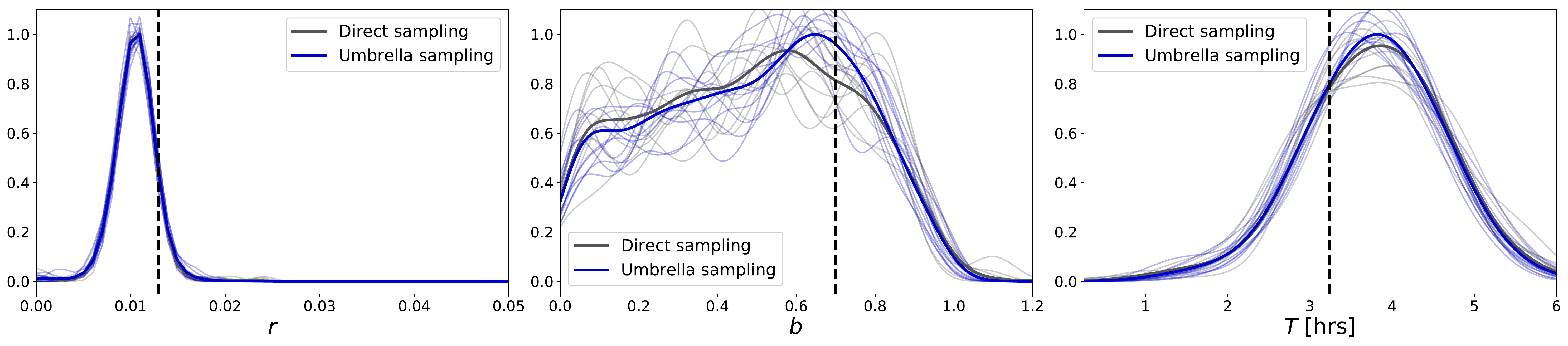}
    \caption{Posterior distributions of $r$, $b$, and $T$ for a simulated non-grazing transit of a super-Earth orbiting a K-dwarf star (simulation SE). See Table \ref{tab:sim_parameters} for simulated model parameters and Figure \ref{fig:simulated_photometry_valley} for the simulated photometry. Each thin line represents a 2000 sample chain from a single independent Monte Carlo run, while the thick lines give the combined results of 20 such runs. Vertical dashed lines represent ground-truth parameter values. The transit trajectory is far from grazing ($b=0.7, r=0.012$), and the two methods produce comparable results ,as expected.}
    \label{fig:posteriors_SE}
\end{figure*}

Posterior distributions for the super-Earth simulation produce consistent results regardless of which method is used (Figure \ref{fig:posteriors_SE}). Because the transit trajectory is far from grazing ($b=0.7, r=0.012$) this agreement is to be expected. The marginalized constraints for this case are $r_p = 1.06 \pm 0.17 R_{\oplus}$, $b = 0.51 \pm 0.30$, correctly identifying the planet as a rocky object with a non-zero, non-grazing impact parameter.

Once again, the primary advantage of umbrella sampling is that it affords us confidence in our results. A small fraction of the posterior samples are consistent with $b > 1$, and by employing umbrella sampling we can be sure that we have correctly weighted the high-$b$ tail of the distribution, whereas with direct sampling alone there would be ambiguity as to whether the tail has been properly explored. In this case, direct sampling does manage to produce the correct result, but we only know this because we have also fit the transit using umbrella sampling. In this specific case, a larger fraction of samples consistent with a grazing trajectory would have made the radius uncertainty larger, which in turn would make the composition of the planet ambiguous, a major detriment for studies of planets near the radius valley.

\subsubsection{Simulation MN: A barely grazing mini-Neptune}

\begin{figure*}
    \centering
    \includegraphics[width=0.9\textwidth]{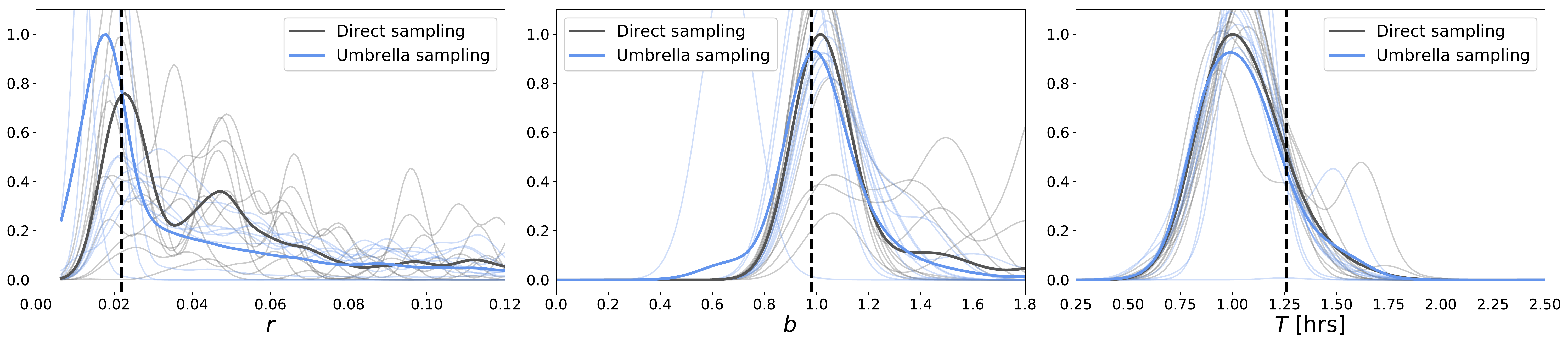}
    \caption{Posterior distributions of $r$, $b$, and $T$ for a simulated barely grazing transit of a mini-Neptune orbiting a K-dwarf star (simulation MN). See Table \ref{tab:sim_parameters} for simulated model parameters and Figure \ref{fig:simulated_photometry_valley} for the simulated photometry. Each thin line represents a 2000 sample chain from a single independent Monte Carlo run, while the thick lines give the combined results of 20 such runs. Vertical dashed lines represent ground-truth parameter values. In this case, umbrella sampling produces obviously improved results, as direct sampling struggles to smoothly explore the grazing regime.}
    \label{fig:posteriors_MN}
\end{figure*}

For our mini-Neptune simulation, posterior inferences made via umbrella sampling are significantly better than those made via direct sampling (Figure \ref{fig:posteriors_MN}). Whereas umbrella sampling returns $r_p = 2.17^{+6.16}_{-0.55}\ R_{\oplus}$, $b = 0.96^{+0.10}_{-0.02}$, direct sampling returns $r_p = 5.3^{+19.4}_{-2.95}\ R_{\oplus}$, $b = 1.03^{+0.19}_{-0.04}$. The reduced precision in $r_p$ from direct sampling will have dramatic consequences for understanding the composition of the individual planet. Even though there is indeed a fairly large uncertainty on the planet radius no matter what method is used - which is to be expected for grazing transits - the implied planet composition is far more ambiguous using direct sampling.

\subsection{A rocky planet in the M-dwarf habitable zone}\label{subsec:case_study_MHZ}

\begin{figure}
    \centering
    \includegraphics[width=0.45\textwidth]{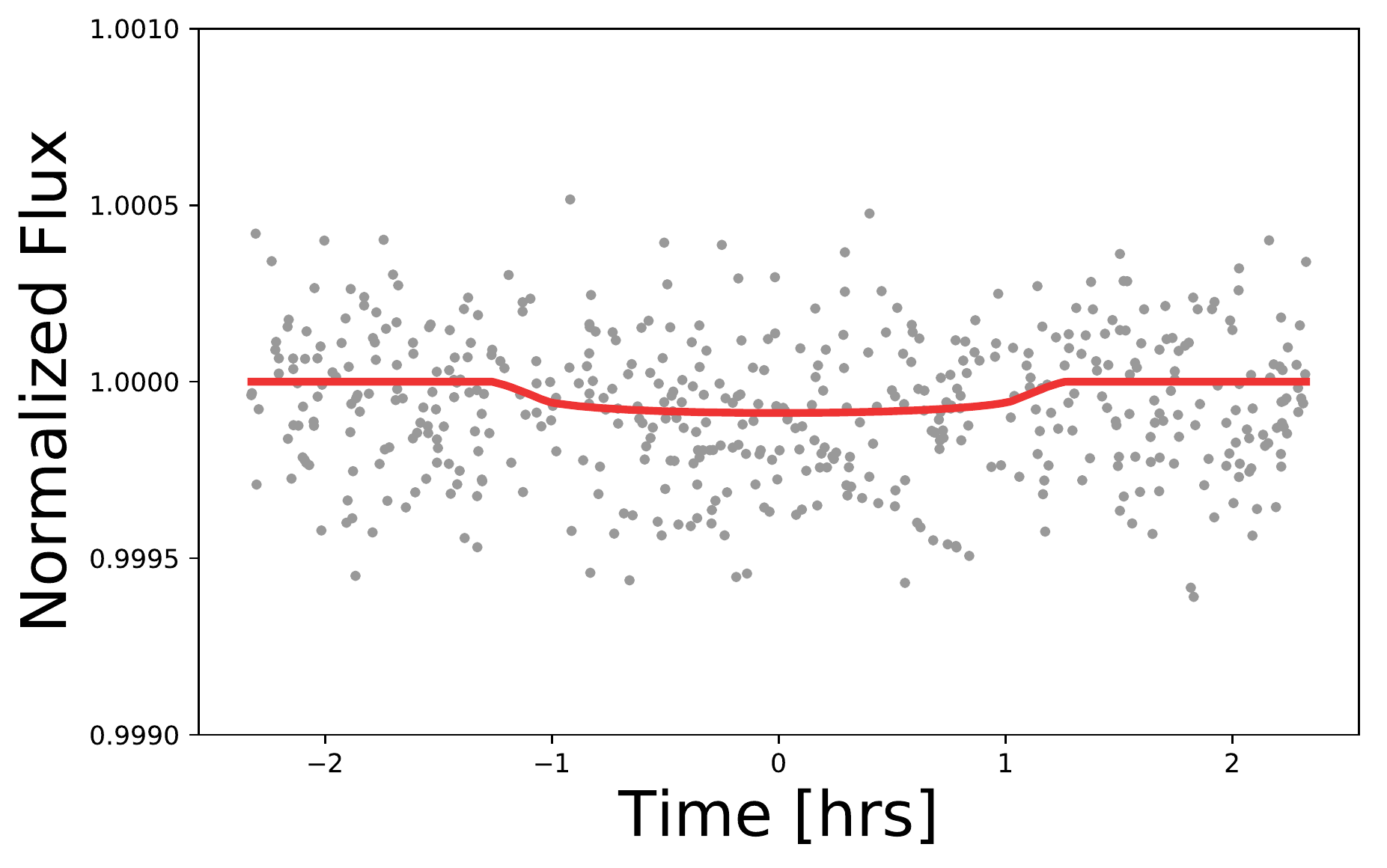}
    \caption{Simulated photometry for a Mercury radius planet orbiting in the habitable zone of an M-dwarf host star (model MHZ). Ground truth simulation parameters are collected in Table \ref{tab:sim_parameters}. See \S\ref{subsec:case_study_MHZ} for discussion.}
    \label{fig:simulated_photometry_mhz}
\end{figure}

For our final test (simulation MHZ), we place a Mercury-sized planet ($r_p = 0.38 R_{\oplus}$) on a $P=37$ day orbit around a $R_{\star} = 0.38 R_{\odot}$ M dwarf, which puts the planet squarely in that star's habitable zone. See Figure \ref{fig:simulated_photometry_mhz} for the simulated photometry and Table \ref{tab:sim_parameters} for the ground truth simulation parameters.

\begin{figure*}
    \centering
    \includegraphics[width=0.9\textwidth]{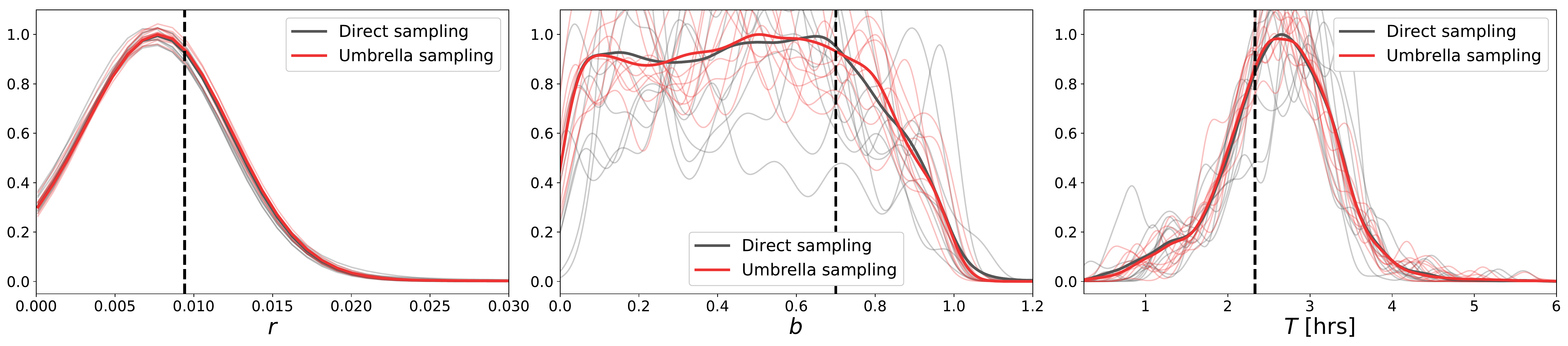}
    \caption{Posterior distributions of $r$, $b$, and $T$ for a simulated barely grazing transit of a Mercury-sized planet orbiting in the habitable zone of an M-dwarf star (simulation MHZ). See Table \ref{tab:sim_parameters} for simulated model parameters and Figure \ref{fig:simulated_photometry_mhz} for the simulated photometry. Each thin line represents a 2000 sample chain from a single independent Monte Carlo run, while the thick lines give the combined results of 100 such runs. Vertical dashed lines represent ground-truth parameter values. For this test case, both direct sampling and umbrella sampling perform equally well.}
    \label{fig:posteriors_MHZ}
\end{figure*}

We find that direct sampling and umbrella sampling perform equally well for this test case, with both methods recovering the true values for $r$, $b$, and $T$ with nearly identical accuracy (Figure \ref{fig:posteriors_MHZ}). Specifically, both methods find $r_p = 0.32 \pm 0.20\ R_{\oplus}$ and a broad, predominantly non-grazing distribution for $b$. Yet even in this case where marginalized statistics are nearly identical, umbrella sampling still confers an advantage over direct sampling. Because the posterior distribution for impact parameter extends above $b = 1$ for both methods, with direct sampling we cannot be certain that the entire full posterior space has been adequately explored. Rather, it is possible we encountered the usual bottleneck at the grazing/non-grazing boundary, leaving the grazing regime undersampled. With umbrella sampling, however, we can be confident \textemdash without the need for follow-up observations \textemdash that the posterior geometry has been fully explored, meaning that the planet is indeed on a non-grazing orbit and therefore has an accurately measured radius.

\section{Analysis of real targets}\label{sec:real_systems}

We will now use umbrella sampling to estimate the impact parameters and radii of several KOI planet candidates with $b > 1$ reported by the NASA Exoplanet Archive cumulative KOI table. For the KOIs in multiplanet systems, we also fit the sibling planets (which are not on grazing trajectories) in order to verify that our results are consistent with previous measurements. For the KOIs in single planet systems, we select an unrelated planet with with a non-grazing $b$ but otherwise similar properties to serve as a basis of comparison.

Of the four $b>1$ candidates we investigate, we find that two are almost certainly on non-grazing trajectories (in conflict with reported values) and two are unambiguously on grazing trajectories  (in agreement with reported values). For the two now non-grazing targets, we report updated values on $r_p$ and $b$ which substantially revise our interpretation of these objects' properties. For the two grazing targets, our results are in agreement with previous measurements and thus place a broad upper limit on $r_p$ and $b$. The advantage of umbrella sampling - even when results are unchanged from literature values - is that because we have explored the full range of geometries, we can be confident that our results are reliable and not artifacts of sampler inefficiencies.

Our data reduction and transit fitting pipeline is described below. Similarly to our experiments with simulated data, we have endeavored to use standard techniques wherever possible, except of course for the steps of the procedure which directly implement umbrella sampling.

We begin by downloading the Pre-search Data Conditioning Simple Aperature Photometry (PDCSAP) flux from the Mikulski Archive for Space Telescopes (MAST). We then flag bad cadences and remove any large outliers with iterative sigma clipping at the $5\sigma$ level. We next remove long-term trends using a Gaussian Process (GP) implemented by \texttt{celerite} \citep{ForemanMackey2017}. For the GP kernel, we adopted a stochastically driven simple harmonic oscillation SHOTerm\footnote{\url{https://celerite.readthedocs.io/}}, which has been shown to produce good results for astronomical time series \citep{ForemanMackey2017}. In order to protect the transit shape during detrending, we mask all cadences within 1.5 transit durations of each expected mid-transit time and project our GP trend across the masked transit region.

To account for possible transit timing variations (TTVs), we read in the transit time measurements of \citet{Holczer2016} and fit a smooth model to these using a GP regression and a Matern-3/2 kernel. We obtain a self-consistent starting estimate for the transit shape and transit times by first fitting $\{P, t_0, \ln r, b, T\}$ and holding transit times fixed, then reversing the procedure to hold transit shape parameters fixed and fitting independent transit times. Finally, we model the independent transit times using a 1st-3rd order polynomial and either zero or one single frequency sinusoids. The complexity of the TTV model was selected based on the Akaike Information Criterion \citep[AIC;][]{Akaike1974}.  For all steps of this TTVs initialization procedure we hold limb darkening coefficients fixed to the theoretical values obtained from the NASA Exoplanet Archive.

While sampling from the posterior, we hold transit times fixed at our low order polynomial + sinusoid model and sample each of the umbrellas independently following the prescription in \S\ref{sec:umbrella} and \S\ref{sec:sampler_comparison}. This means that the free parameters in the model are $\{\ln T, q_1, q_2, F, \ln\sigma_F\}$, plus either $\{\ln r,b\}$ for the N umbrella, $\{\ln r, \gamma\}$ for the T umbrella, or $\{\ln\lambda, \gamma\}$ for the G umbrella. As usual, we adopt a Rayleigh prior for the eccentricity, with scale parameter, $\sigma_e$, chosen to match the architecture of the system under consideration (see below). Stellar density priors were taken from the \textit{Gaia-Kepler} catalog \citep{GaiaDR2, Berger2018}.

For simplicity, we only consider non-overlapping transits and fit planets one at a time for multiplanet systems. Overlapping transits were defined as any transit pair for which $|t_{0,b}-t_{0,c}| < (T_b + T_c)$ for any two planets b and c. Each HMC run consisted of two independent chains, with each chain run for a default length of 10,000 tuning steps and 5,000 draws, generating 10,000 samples total per run. In a few cases, the chains did not converge on our first attempt to fit the data, in which case extending the length of the tuning phase remedied the issue.

After drawing samples from all three windows - N, T, and G - we check that the posterior samples of $r$, $b$, and $T$ are consistent between the sub-distributions $\pi_N$, $\pi_T$, and $\pi_G$ for each planet. This does not mean that the distributions must overlap completely (indeed, they are expected not to), but rather that they have at least some overlap, with perhaps some modest tension between umbrellas. In practice, we found that posterior sub-distribution were nearly always either obviously consistent or obviously inconsistent, with the later case indicating that the algorithm had not been properly tuned prior to sampling. In some cases, even though sub-distributions were clearly inconsistent when considered simultaneously, results initially appeared reasonable when each umbrella was considered in isolation. Thus, our method provides a new avenue for verifying that the results of a transit fit are trustworthy: if Markov chains do not properly behave within all three windows and produce self-consistent results, we know to investigate further. Thus, the sub-distribution comparison step of our algorithm builds in an extra redundancy for checking convergence.

\subsection{KOI-2068}\label{subsec:KOI-2068}

KOI-2068 is a $0.91 R_{\odot}$ star hosting a single planet candidate at $P=42$ days with $1\sigma$ upper limits $b\leq58$ and $r_p\leq42 R_{\oplus}$. With signal-to-noise $S/N=21$ and a disposition score of $0.89$, the object is unlikely to be a false positive. This combination of degenerate, poorly constrained $r$ and $b$ values plus a low false positive probability makes this object an ideal test case for our umbrella sampling scheme. As a comparison target, we select KOI-2285, a $0.87 R_{\odot}$ star hosting a single confirmed planet at $P=38$ days, with $b=0.26\pm0.23$, $r_p=2.79\pm0.30$, and $S/N=24$. For both targets, we set the Rayleigh eccentricity prior scale at $\sigma_e = 0.21$, the \citet{Mills2019} single-planet value.

\begin{figure*}
    \centering
    \includegraphics[width=0.75\textwidth]{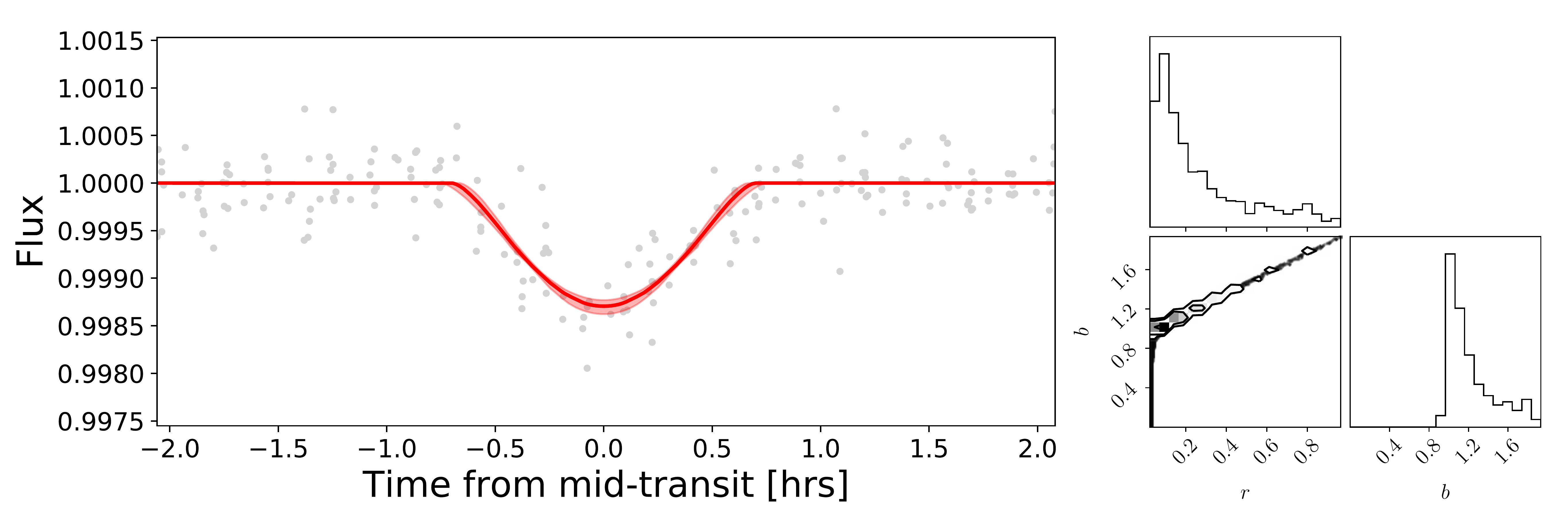}
    \caption{Phase-folded transits of KOI-2068 with posterior model from umbrella sampling overplotted in red. The pink shaded region shows the $1\sigma$ credible interval. The V-shape of the transit is apparent, which is suggestive of a grazing transit. The posterior corner plot for $r$ and $b$ is shown on the right.}
    \label{fig:KOI_2068_transit_posterior}
\end{figure*}

After sampling, for KOI-2068.01 we recover $r_p = 17.7^{+35.9}_{-10.3} R_{\oplus}$, $b=1.13^{+0.37}_{-0.12}$, which is unfortunately not an appreciably different constraint on $r_p$ than the literature value. However, all is not lost. Visual inspection of the posterior transit model (Figure \ref{fig:KOI_2068_transit_posterior}, left panel) suggests that the transit shape is remarkably well constrained, even if the individual parameters are not. Moreover, posterior distributions of $r$ and $b$ (Figure \ref{fig:KOI_2068_transit_posterior}, right panel) are smooth and well behaved, showing a clear preference for grazing geometries. Because umbrella sampling ensures that we have explored the full parameter space, we can now be confident that the weak upper limits on $r$ and $b$ arise from inherent limitations of the data themselves, not from a failure of the sampler. Furthermore, we can now place a confident lower limit on the impact parameter, which will be useful for any follow-up work. A deeper investigation, which is beyond the scope of this work, might reveal transit depth variations or transit duration variations which could be able to place far more precise constraints on $b$ and, consequently, on $r$ \citep{Dawson2020}.

For the comparison target, we recover $r_p = 2.88 \pm 0.15 R_{\oplus}$, $b=0.55 \pm 0.27$, $T = 4.17 \pm 0.16$ hrs, consistent with the literature values. We will not comment on the relative precision or accuracy of our results vs reliable literature results, as any differences in measured values are more likely to be driven by differences in data reduction techniques than by which sampling method was used. The important point is that our analysis was able to reproduce known reliable results, validating our pipeline and affording us confidence in any new measurements which improve upon the state of the art.

\subsection{KOI-2150}\label{subsec:KOI-2150}

KOI-2150 is a $0.94 R_{\odot}$ star hosting two planet candidates, both with impact parameters greater than unity reported on the NASA Exoplanet Archive. The inner planet ($P=19$ days) has $1\sigma$ upper limits $b\leq73$ and $r_p \leq 38 R_{\oplus}$, while the outer planet ($P=45$ days) has $b\leq72$ and $r_p \leq 94 R_{\oplus}$. Both candidates have a disposition score $>0.99$, indicating that neither is likely to be a false positive. We set the Rayleigh eccentricity prior scale at $\sigma_e = 0.0355$, the \citet{Mills2019} multi-planet value.

\begin{figure}
    \centering
    \includegraphics[width=0.45\textwidth]{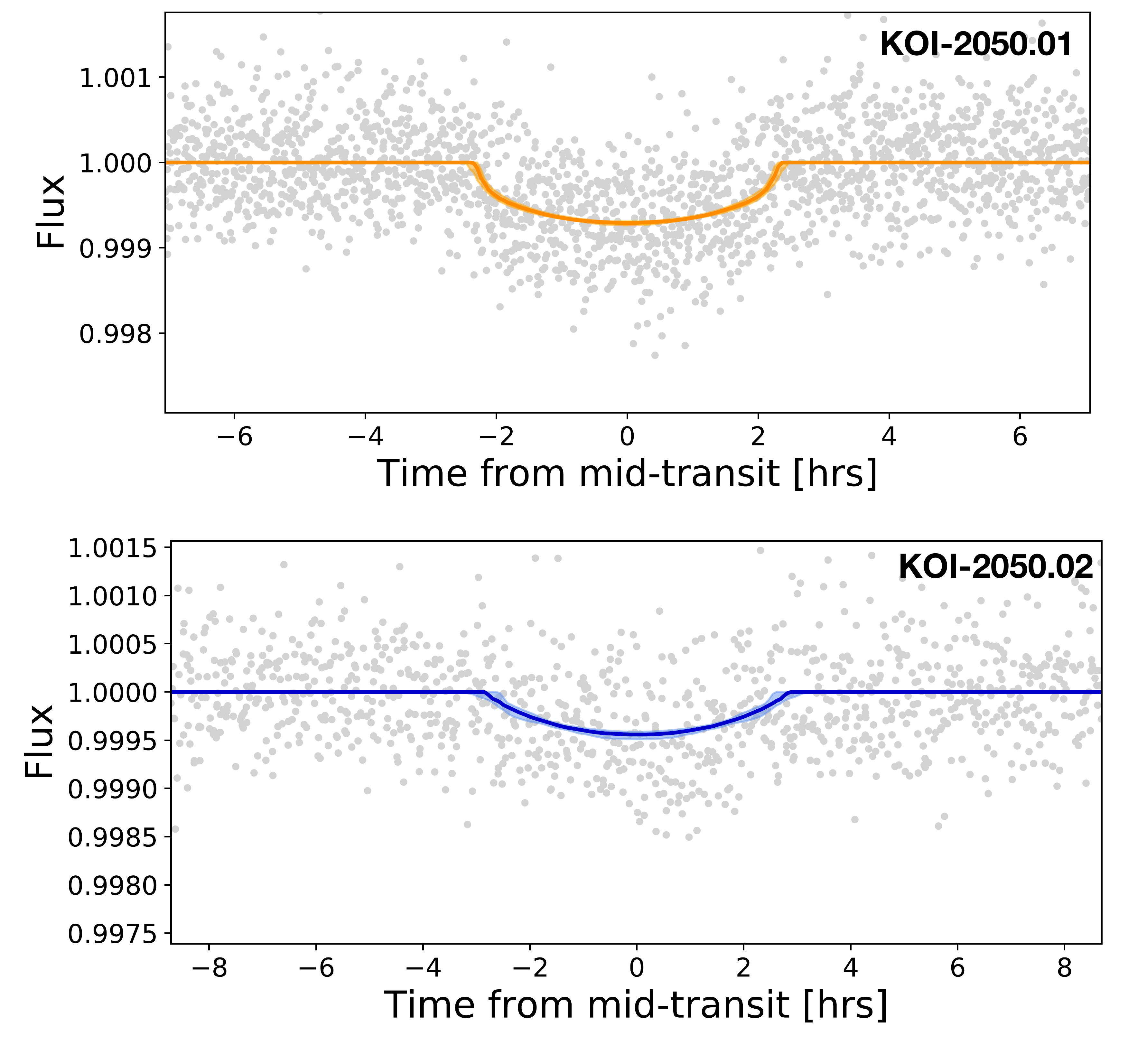}
    \caption{Posterior transit models for the two candidate planets orbiting K01-2050. Shaded regions (just barely visible on the plots) show $1\sigma$ credible intervals.}
    \label{fig:KOI_2150_transit_posterior}
\end{figure}

After umbrella sampling, for the inner planet we recover $b=0.35\pm0.32$, $r_p=2.45\pm0.26 R_{\oplus}$ (11\% radius uncertainty), and for the outer planet we recover $b=0.64^{+0.39}_{-0.43}$, $r_p=2.01^{+5.40}_{-0.26} R_{\oplus}$. Thus, umbrella sampling places both objects on non-grazing trajectories - albeit with poorly constrained impact parameters - and finds a plausible radius for each. In this case, umbrella sampling has significantly outperformed the standard method. These candidates are probably mini-Neptunes, both possessing individual properties consistent with a depleted radius valley \citep{FultonPetigura2018, VanEylen2018} and relative sizes consistent with the ``peas in a pod'' hypothesis \citep{Weiss2018}, adding further credulity to our results. Posterior transit models are shown in Figure \ref{fig:KOI_2150_transit_posterior}.

\subsection{KOI-1426}\label{subsec:KOI-1426}

KOI-1426 is a $0.90 R_{\odot}$ star hosting two confirmed planets and one planet candidate. The two confirmed planets have well-constrained properties reported on the NASA Exopplanet Archive (KOI-1426.01: $P=39$ days, $r_p=2.81\pm0.04 \ R_{\oplus}$, $b=0.03^{+0.33}_{-0.03}$, KOI-1426.02: $P=75$ days, $r_p=6.39\pm0.10 R_{\oplus}$, $b=0.80^{+0.01}_{-0.06}$), but the candidate planet (KOI-1426.03: $P=150$ days, $r_p \leq 36 R_{\oplus}$, $b\leq68$) exhibits the $r-b$ degeneracy. Unlike either candidate in the KOI-2150 system, KOI-1426.03 possesses an impact parameter constraint $b=1.25^{+67}_{-0.17}$ that marks its orbit (if real) as unambiguously grazing. All three planets possess highly significant TTVs with similar periodicities and amplitudes \citep{Holczer2016}, indicating that they are unlikely to be false positives. Consequently, we set the Rayleigh eccentricity prior scale at $\sigma_e = 0.008$, the \citet{LithwickXieWu2012} value.

\begin{figure}
    \centering
    \includegraphics[width=0.45\textwidth]{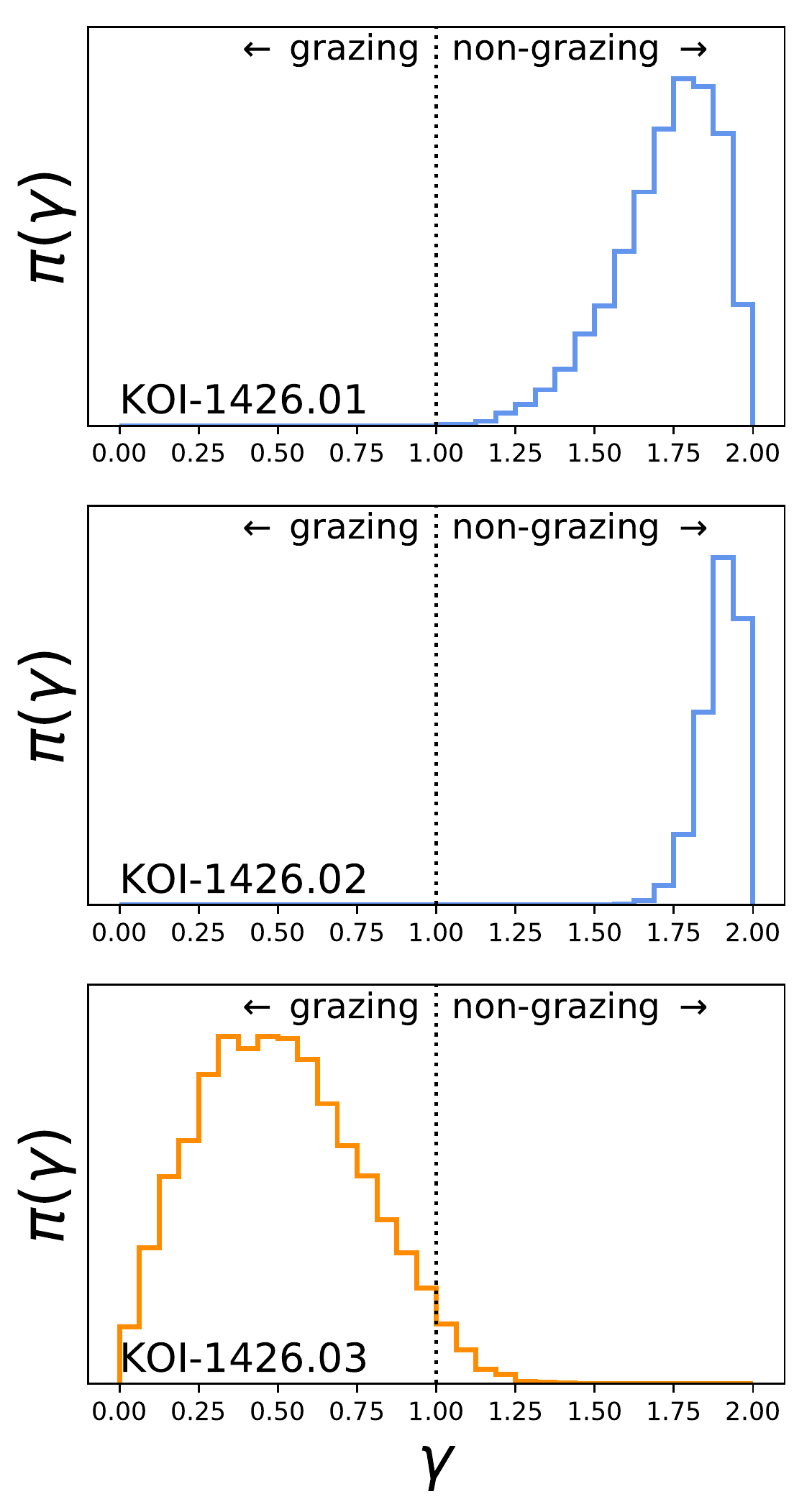}
    \caption{Samples of $\gamma$ obtained under the transition umbrella, $\psi_T$, for the three planets in the KOI-1426 system. Neither of the two confirmed objects (top two panels, blue) have any posterior samples $\gamma < 1$, indicating that both planets are almost certainly on non-grazing trajectories. For these two objects, we can comfortably ignore the grazing umbrella, $\psi_G$, and perform a two-window analysis. The candidate object (bottom panel, orange) has $97\%$ of samples with $\gamma < 1$, indicating that this planet is probably on a grazing trajectory. For this object, we might choose to ignore the non-grazing umbrella, $\psi_N$, but the more conservative approach would be to perform a three-window analysis as usual. The distributions seen here for KOI-1426 are typical of Kepler targets, in that samples of $\gamma$ from the transition umbrella can often be used to rule out/in certain transit geometries.}
    \label{fig:transition_K01426}
\end{figure}

Our umbrella sampling analysis confirms the grazing transit hypothesis for KOI-1426.03, finding $b=1.13^{+0.24}_{-0.12}, r=25.0^{+21.9}_{-9.9} R_{\oplus}$ and $<4\%$ of posterior samples drawn under the transition umbrella consistent with a non-grazing geometry. In contrast, none of the samples drawn from the transition window for either of the two confirmed planets were consistent with a grazing geometry (Figure \ref{fig:transition_K01426}), highlighting the utility of our approach for distinguishing grazing from non-grazing transits.  As expected, our posterior results for the two confirmed planets (KOI-1426.01: $r_p=2.72\pm0.07 \ R_{\oplus}$, $b=0.26\pm0.15$, KOI-1426.02: $r_p=6.52\pm0.16 R_{\oplus}$, $b=0.84\pm0.02$) are consistent with the literature values. Although the uncertainty on the radius of the grazing candidate planet remains high at ${\sim}70\%$, the transit shape is extremely well constrained by the data, reminiscent of the results for KOI-2068.

\begin{figure}
    \centering
    \includegraphics[width=0.45\textwidth]{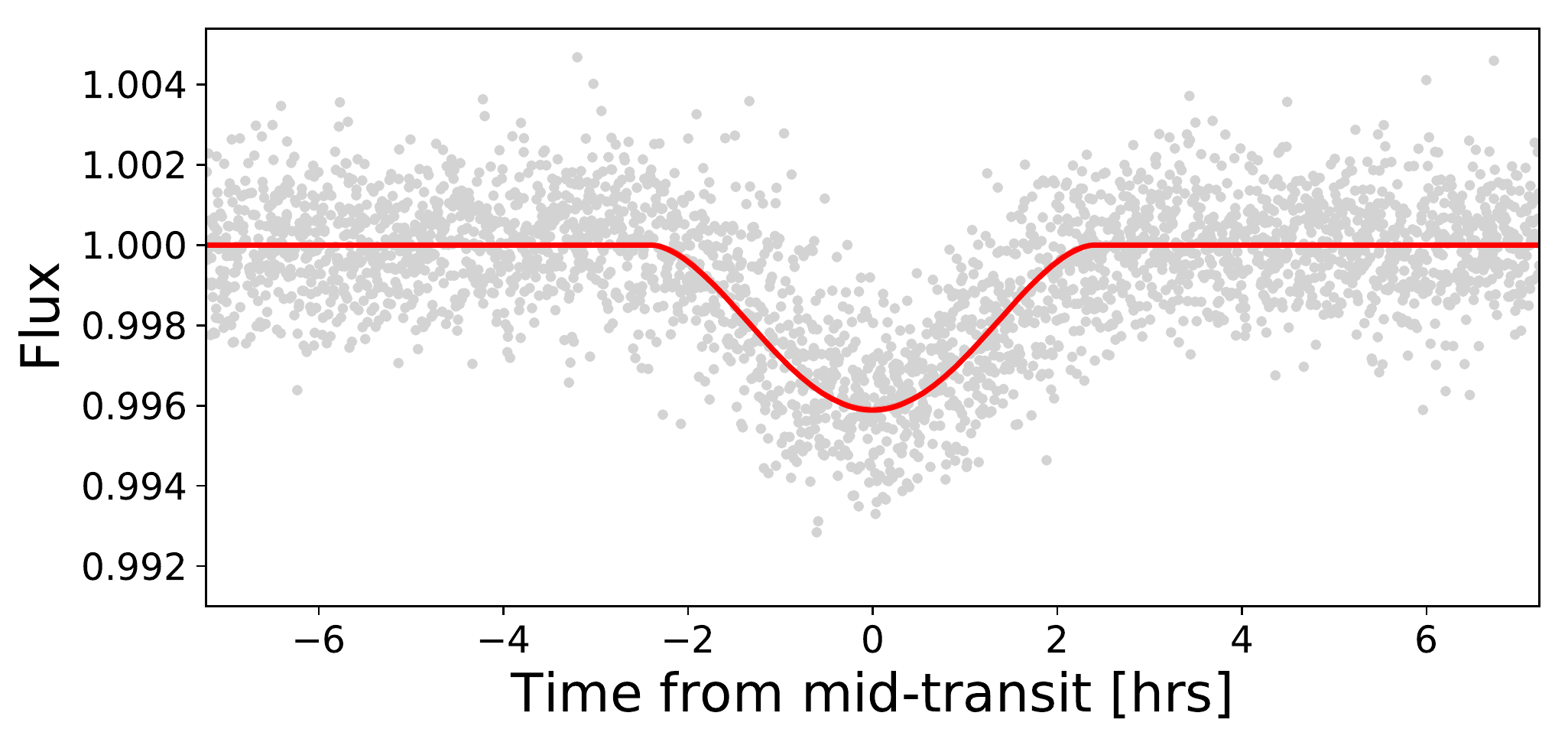}
    \caption{Posterior transit model for K01-1426.03. The $1\sigma$ credible interval is so narrow that it is not even visible on the plot.}
    \label{fig:KOI_1426_transit_posterior}
\end{figure}

Revealing the true properties of the KOI-1426 system will likely require a full photodynamical analysis. In retrospect, this is unsurprising for two reasons. First, all three planets exhibit large transit timing variations \citep{DiamondLowe2015, Holczer2016} which may be insufficiently characterized by our parametric model. Second, grazing transits often have a time-dependent transit shape \citep{Hamann2019, Dawson2020}, and thus our approximation of an invariant transit shape may yield biased inferences. These complications do not mean that our present efforts to model the system were a waste of time. On the contrary, the results obtained with umbrella sampling will serve as useful priors for setting up the computationally expensive photodynamical model. Informed priors (such as the fact that KOI-1426.02 is both real and on a grazing trajectory) can place meaningful limits on the system architecture and thereby greatly improve both the accuracy and efficiency of the full photodynamical treatment. Furthermore, the techniques of photodynamics and umbrella sampling are not mutually exclusive, and it may ultimately prove necessary to combine the two methods in order to achieve a definitive result for this or other dynamically active systems.

\section{Summary and recommendations}\label{sec:summary}

We have introduced a new method for modeling exoplanet transit lightcurves which explicitly accounts for the differences in transit geometry between grazing and non-grazing trajectories. Our technique employs the well-established framework of umbrella sampling \citep{TorrieValleau1977} by splitting the transit fitting problem into three sub-problems, each restricted to either the grazing, non-grazing, or transition regions of the parameter space. We draw samples independently from each window using an MCMC sampler to produce three posterior sub-distributions which we then recombine into a single joint posterior distribution using the Eigenvecor Method of Umbrella Sampling \citep{Thiede2016, Dinner2017}.

Although umbrella sampling is widely used by molecular dynamicists and biochemists, it has only recently begun to gain the attention of astronomers \citep{Matthews2018}. Yet umbrella sampling is itself a general statistical tool not tied to any particular content domain. At heart, umbrella sampling is designed to estimate complicated posterior geometries (e.g. isolated modes or degeneracy ridges) - geometries of the sort that arise frequently in astrophysical studies. By applying umbrella sampling to a familiar astronomical problem and illustrating its efficacy, we hope to raise awareness of this powerful statistical technique which is well suited to astronomical data analysis. To aide astronomers first learning to use umbrella sampling, we have provided an introductory Python tutorial at \url{https://gjgilbert.github.io/tutorials/umbrella_sampling/}.

Our umbrella sampling routine reliably produces posterior estimates of planetary radii and impact parameters which are more accurate than estimates obtained using a standard approach. We tested our method under a wide range of conditions using both real and synthetic data, finding that umbrella sampling performed at least as well as \textemdash and usually better than \textemdash the standard direct sampling approach for every star-planet configuration we considered. Moreover, even in cases where umbrella sampling did not provide higher precision estimates than direct sampling, we were able to have greater confidence in the results of umbrella sampling because only this method is able to efficiently explore deep in the the grazing regime.

Throughout this paper, we have offered numerous suggestions for how to modify existing transit modeling procedures in order to produce more robust posterior estimates. We now summarize these recommendation here.

\begin{enumerate}
    \item Before fitting any transit model, perform an exploratory analysis restricted to the region of parameter space immediately surrounding the grazing/non-grazing transition at $r=1-b$. This exploration can be efficiently executed using our $\{r,\gamma\}$ basis and and the transition umbrella, $\psi_T$.
    \item If all samples $\pi_T(\gamma)$ are consistent with a non-grazing geometry (i.e. all $\gamma > 1$), one may proceed with a standard analysis, restricting the model to non-grazing geometries. Conversely, if all samples $\pi_T(\gamma)$ are consistent solely with a grazing geometry (all $\gamma < 1$), one may instead restrict the model to grazing geometries and sample using our new $\{\lambda,\gamma\}$ basis.
    \item If, however, samples $\pi_T(\gamma)$ are mixed between grazing and non-grazing geometries, the transit should be modeled using the scheme we have described in detail in \S\ref{subsec:basis_specification} and \S\ref{subsec:full_model}. 
    \item After sampling from under the various umbrellas $\psi_i$, compare posterior sub-distributions $\pi_i$ for each transit parameter to ensure that inferences are consistent between samples drawn from different windows. If samples are in disagreement, closer investigation is needed. This comparison step provides an additional convergence check for the user.
    \item For planets inferred to orbit on a grazing trajectory, consider whether a fully photodynamical analysis is needed. If so, the results obtained via umbrella sampling will serve as useful priors for initializing the more computationally expensive photodynamical model, thereby improving efficiency.
\end{enumerate}

For simplicity throughout this work, we always used the full first-to-fourth contact transit duration $T_{14}$ as our transit duration because it is defined regardless of transit geometry. However, the center-to-center duration (1.5 to 3.5 contact), $T_{c-c}$ is often better constrained by the data and is therefore often preferred as a basis parameter as long as it is defined, which it will be as long as $\gamma > 0$. Given our window bounds (Equations \ref{eq:psi_N}-\ref{eq:psi_G}), this means that we are free to use $T_{c-c}$ in place of $T_{14}$ for the N and T umbrellas. Swapping one $T$ for another adds an additional step to the procedure, in that a consistent $T$ must be used to produce the final joint posterior distribution. Fortunately, once samples have been obtained for all parameters, determining $T_{14}$ from $T_{c-c}$ is a matter of straightforward arithmetic, and vice versa. One may alternatively use the full-width-half-max transit duration, $T_{\rm FWHM}$, which can be defined in relation to the transit depth, $\delta$, even for strongly grazing transits. The downside of using $T_{\rm FWHM}$ is that it is numerically more difficult to determine and introduces a new covariance between $T$ and $\delta$. When applying umbrella sampling in the future, $T_{c-c}$ should probably be adopted whenever possible.

For most of the history of exoplanet science, uncertainties on planetary radii and orbital parameters have been dominated by uncertainties on stellar parameters. Now, however, with improved stellar radius estimates from Gaia \citep{GaiaDR2}, and high resolution spectroscopy \citep{Johnson2017, Petigura2017}, the details of the transit fitting problem have once again become relevant for obtaining state-of-the-art estimates of planet properties. Statistical studies of exoplanets will remain dominated by the population of transiting planets for at least the next decade, and so transit modeling will remain at the foundation of many astrophysical analyses. By adopting umbrella sampling as a new tool, we will ensure that our understanding of exoplanet demographics, architectures, and formation histories will reach as far as the data allow.

\acknowledgements
GJG is supported by a NASA Future Investigators in Earth and Space Sciences and Technology (FINESST) Felloship, Grant Number 80NSSC20K1533.

This study made use of data products from the \Kepler mission hosted on the NASA Exoplanet Archive. Some of the data were obtained from the Mikulski Archive for Space Telescopes (MAST) at the Space Telescope Science Institute. These data can be accessed via \dataset[10.17909/T98304]{\doi{10.17909/T98304}}. This study also made use of computational resources provided by the University of Chicago Research Computing Center.

We thank the anonymous referee for their detailed feedback and many helpful comments which greatly improved the quality of this manuscript. We thank Andrey Kravstov, Dan Fabrycky, Leslie Rogers, Fred Ciesla, Erik Petigura, Mason MacDougall, Dan Foreman-Mackey, and Louis Smith for thoughtful conversations which guided the direction of this study.

Software: \texttt{astropy} \citep{astropy:2013, astropy:2018}, 
\texttt{celerite} \citep{ForemanMackey2017} \texttt{exoplanet} \citep{ForemanMackey2021}, \texttt{numpy} \citep{numpy:2020}, \texttt{PyMC3} \citep{pymc3:2016}, \texttt{scikit-learn} \citep{scikit-learn:2011}, \texttt{scipy} \citep{scipy:2020}, \texttt{starry} \citep{Luger2019}, \texttt{usample} \citep{Matthews2018}

\bibliography{ms} \bibliographystyle{apj}

\appendix

\section{A. Derivation of lambda}\label{appx:A}

For a uniform surface brightness (i.e. zero limb darkening) stellar source, the fractional change in flux during transit, as derived by \citet{MandelAgol2002}, is

\begin{equation}\label{eq:mandel_agol}
    \Lambda(r,z) =
    \begin{cases}
			\frac{1}{\pi}\big[ r^2\kappa_0 + \kappa_1 - \frac{1}{2}\zeta \big] & \text{$1-r < z < 1+r$}\\
            r^2 & \text{$0 \leq z \leq 1-r$}
    \end{cases}
\end{equation}

where $r\equiv r_p/R_{\star}$ is the planet-to-star radius ratio, $z = z(t)$ is the is the normalized separation of centers, and

\begin{equation}
\begin{aligned}
    &\kappa_0 = \cos^{-1}[(r^2+z^2-1)/2rz] \\
    &\kappa_1 = \cos^{-1}[(1-r^2+z^2)/2z] \\
    &\:\;\zeta = \sqrt{4z^2 - (1+z^2-r^2)^2}
\end{aligned}
\end{equation}

Making the substitution $z(t_0) \rightarrow b$ gives the area of overlap at midtransit, $\Lambda_0$, which is equivalent to the transit depth, $\delta$, as long as we continue to neglect limb darkening. For non-grazing trajectories, $\Lambda_0$ is independent of $b$, whereas for grazing ones $\Lambda_0$ has a complicated dependence on $b$ and $r$, which become highly degenerate. It is this complicated geometry that leads to considerable difficulties when modeling grazing transits.

At the grazing/non-grazing transition boundary, $b=1-r$ and $\Lambda_0 = r^2$; at the grazing limit where the planet just barely transits, $b=1+r$ and $\Lambda_0 = 0$. If we define the convenience variable $\beta \equiv 1 - b$, then the two endpoints of grazing geometries in $(\beta,\Lambda_0)$ space for a given $r$ are $(r,r^2)$ and $(-r,0)$. The equation of a straight line between these two points is

\begin{equation}
    \Lambda_0 = \Big(\frac{0-r^2}{-r-r}\Big)(\beta + r) = \frac{r}{2}(\beta + r) = \frac{r}{2}\beta + \frac{r^2}{2}
\end{equation}

Defining $\lambda \equiv 2\Lambda_0$ gives $\lambda = \beta r + r^2$, which is the definition of $\lambda$ we presented in Equation \ref{eq:lam_gam} in the main text.

\section{B. Accounting for implicit priors}\label{appx:B}

Our philosophy throughout this work is to place minimally informative priors on all variables wherever possible. For $r$ and $b$, there are two common approaches to do so. The first is to assume that $r$ and $b$ are uncorrelated and then then draw ($r, b$) pairs uniformly from the physically permissible region of the $r-b$ plane \citep{Espinoza2018}. This prior choice is sensible, but introduces a marginal prior on $r$. The second approach, which we opt to use here, is to establish a joint ($r,b$) distribution which produces a flat marginal prior on both $r$ and $b$. This latter approach is the default \texttt{ImpactParameter} distribution implemented in the popular Python package \texttt{exoplanet} \citep{ForemanMackey2021}.

\medskip
Ideally, proposal distributions, $p(\theta)$, will match the desired prior distributions, making the analysis uncomplicated. In situations where specifying the desired prior in closed-form is impossible or inconvenient, a useful strategy is to write down a simple proposal distribution and then modify the likelihood function, $\mathcal{L}$, such that the desired prior is produced.

\medskip
For a set of observed data points, $\boldsymbol{y}$, a corresponding model, $\boldsymbol{\hat y} = f(\boldsymbol{\theta})$, and assuming Gaussian residuals, the log-likelihood functions is

\begin{equation}
    \ln\mathcal{L}(\boldsymbol{\theta} | \boldsymbol{y}) = -n\ln\big(\sigma\sqrt{2\pi}\big) - \sum_i^n \frac{1}{2\sigma^2}\big(y_i - \hat y_i\big)^2
\end{equation}

which we adopt here.

\medskip
Addressing the transit problem specifically, we draw $r$ and $b$ sequentially from conditional proposal distributions

\begin{equation}
\begin{aligned}
    & p(r) \sim \mathcal{U}(r_{\rm min}, r_{\rm max}) \\
    & p(b|r) \sim \mathcal{U}(0, 1+r)
\end{aligned}
\end{equation}

where $\mathcal{U}$ is a uniform interval. The \citet{Espinoza2018} prior can then be recovered by adding $\ln(r+r^2)$ to $\ln\mathcal{L}$, or the \texttt{exoplanet} prior can be produced by adding $\ln(1+r)$. This ``modify-the-likelihood'' approach is admittedly a bit circuitous when applied here, but proves convenient after splitting the problem into various umbrellas, each with their own parameter basis.

\medskip
For the non-grazing (N) umbrella, we draw $r$ and $b$ sequentially as

\begin{equation}
\begin{aligned}
    & p(r) \sim \mathcal{U}(r_{\rm min}, r_{\rm max}) \\
    & p(b|r) \sim \mathcal{U}(0, 1-r)
\end{aligned}
\end{equation}

\noindent Here, the \citet{Espinoza2018} prior can be recovered by adding $\ln(1-r)$ to $\mathcal{L}$, and the \texttt{exoplanet} prior can be produced from the \citet{Espinoza2018} prior by subsequently subtracting $\ln(1+r)$.

\medskip
For the other two umbrellas, we must also account for any implicit priors induced by coordinate transformations. This may be accomplished by adding the log-determinant of the Jacobian, $\ln|J|$ to the log-likelihood. Recall that our two new basis parameters are defined as

\begin{equation}
\begin{aligned}
    &\lambda = r^2 + \beta r\\
    &\gamma = \frac{\beta}{r}
\end{aligned}
\end{equation}

where $\beta \equiv 1-b$.

\medskip
For the transition (T) umbrella, the Jacobian of the coordinate transformation $\{r,b\} \rightarrow \{r, \gamma\}$ is

\begin{gather}
    J_T = 
    \begin{vmatrix}
        \frac{\partial r}{\partial r}  &  \frac{\partial r}{\partial b}      \\
        \frac{\partial \gamma}{\partial r}  &  \frac{\partial \gamma}{\partial b}
    \end{vmatrix}
    = 
    \begin{vmatrix}
        1  &  0      \\
        (b-1)/r^2  &  -1/r      
    \end{vmatrix} 
\end{gather}

so $\ln|1/r|$ should be added to $\ln\mathcal{L}$. We draw ($r, \gamma$) sequentially as

\begin{equation}
\begin{aligned}
    & p(r) \sim \mathcal{U}(r_{\rm min}, r_{\rm max}) \\
    & p(\gamma) \sim \mathcal{U}(0,\gamma')
\end{aligned}
\end{equation}

where $\gamma' = 2$ if $r<0.5$ and $\gamma' = 1/r$ if $r \geq 0.5$; the variable bound on $\gamma$ prevents negative impact parameters from being drawn. When $r < 0.5$, adding $2\ln 2r$ to $\ln\mathcal{L}$ reproduces the \citet{Espinoza2018} prior, and when $r \geq 0.5$ adding $\ln r$ does so. Subsequently subtracting $\ln(1+r)$ again reproduces the \texttt{exoplanet} prior.

\medskip
For the grazing (G) umbrella, the Jacobian of the coordinate transformation $\{r,b\} \rightarrow \{\lambda, \gamma\}$ is

\begin{gather}
    J = 
    \begin{vmatrix}
        \frac{\partial \lambda}{\partial r}  &  \frac{\partial \lambda}{\partial b}      \\
        \frac{\partial \gamma}{\partial r}  &  \frac{\partial \gamma}{\partial b}
    \end{vmatrix}
    = 
    \begin{vmatrix}
        2r + 1 - b  &  -r      \\
        (b-1)/r^2  &  -1/r      
    \end{vmatrix} 
\end{gather}

so $\ln|2 + 2(1-b)/r|$ = $\ln|2 + 2\gamma|$ should be added to $\ln\mathcal{L}$. We draw ($\lambda,\gamma$) sequentially as

\begin{equation}
\begin{aligned}
    & f(\gamma) \sim \mathcal{U}(-1,1) \\
    & f(\lambda|\gamma) \sim \mathcal{U}\Big((\gamma + 1)r_{\rm min}^2, (\gamma+1)r_{\rm max}^2\Big)
\end{aligned}
\end{equation}

Coincidentally, the Jacobian element exactly matches the likelihood adjustment needed to recover the \citet{Espinoza2018} prior, i.e. subtracting $\ln|2 + 2\gamma|$ from $\mathcal{L}$ (or equivalently, omitting the Jacobian term) will produce uniform samples in the $r-b$ plane. Subsequently subtracting $\ln(1+r)$ produces the \texttt{exoplanet} prior.

\medskip
The equations above present a method for establishing desirable prior and proposal distributions on $r$, $b$, $\lambda$, and $\gamma$. In practice, it is usually preferable to sample in $\ln r$ and $\ln \lambda$ in order to allow for a wide range of transit depths (this is indeed the approach taken throughout this work). In this case, the logic of the coordinate transformations remains the same, but the likelihood correction terms are different.

\medskip
For the non-grazing (N) umbrella, we draw $\ln r$ and $b$ sequentially as

\begin{equation}
\begin{aligned}
    & p(\ln r) \sim \mathcal{U}(\ln r_{\rm min}, \ln r_{\rm max}) \\
    & p(b|r) \sim \mathcal{U}(0, 1-r)
\end{aligned}
\end{equation}

Adding $\ln(1-r) + \ln r$ to $\ln \mathcal{L}$ reproduces the \citet{Espinoza2018} prior (and absorbs the Jacobian element). Subsequently subtracting $\ln(1+r)$ reproduces the \texttt{exoplanet} prior. For both cases, the marginal prior on $r$ can be made to be log-uniform be also subtracting $\ln r$ from $\ln \mathcal{L}$. Note that some of these likelihood correction terms cancel, but we have listed them all explicitly for clarity.

\medskip
For the transition (T) umbrella, we draw ($\ln r, \gamma$) sequentially as

\begin{equation}
\begin{aligned}
    & p(\ln r) \sim \mathcal{U}(\ln r_{\rm min}, \ln r_{\rm max}) \\
    & p(\gamma) \sim \mathcal{U}(0,\gamma')
\end{aligned}
\end{equation}

where as before $\gamma' = 2$ if $r<0.5$ and $\gamma' = 1/r$ if $r \geq 0.5$, preventing negative impact parameters from being drawn. When $r < 0.5$, adding $\ln r + \ln 2r$ to $\ln\mathcal{L}$ reproduces the \citet{Espinoza2018} prior (and absorbs the Jacobian element), and when $r \geq 0.5$ adding $\ln r$ does so. Subsequently subtracting $\ln(1+r)$ reproduces the \texttt{exoplanet} prior. For both cases, the marginal prior on $r$ can be made to be log-uniform be also subtracting $\ln r$ from $\ln \mathcal{L}$. As with the N umbrella, we have listed all terms explicitly for clarity, even when they cancel.

\medskip
For the grazing (G) umbrella, we draw ($\ln \lambda, \gamma$) sequentially as

\begin{equation}
\begin{aligned}
    & f(\gamma) \sim \mathcal{U}(-1,1) \\
    & f(\ln\lambda|\gamma) \sim \mathcal{U}\Big(\ln [(\gamma + 1)r_{\rm min}^2], \ln [(\gamma+1)r_{\rm max}^2]\Big)
\end{aligned}
\end{equation}

Adding $2 \ln r$ to $\ln \mathcal{L}$ reproduces the \citet{Espinoza2018} prior (and absorbs the Jacobian element). Subsequently subtracting $\ln(1+r)$ reproduces the \texttt{exoplanet} prior. For both cases, the marginal prior on $r$ can be made to be log-uniform be also subtracting $\ln r$ from $\ln \mathcal{L}$. As with the N and T umbrellas, we have listed all terms explicitly for clarity, even when they cancel.

\end{document}